\documentclass[a4paper,11pt]{article}
\pdfoutput=1
\usepackage{jcappub}
\usepackage{subcaption}
\graphicspath{{plots/}}
\newcommand{\gs}{g_\text{eff}}
\newcommand{\gss}{h_\text{eff}}
\newcommand{\Trh}{T_\text{rh}}
\newcommand{\Tmax}{T_\text{max}}

\newcommand{\rR}{\rho_R}
\newcommand{\rp}{\rho_\phi}
\newcommand{\Gp}{\Gamma_\phi}

\newcommand{\MO}{\texttt{micrOMEGAs}}

\usepackage[normalem]{ulem}

\title{micrOMEGAs 7:\\Beyond standard cosmology} 

\author[a]{G.~Bélanger,}
\author[b]{A.~Belyaev,}
\author[c]{N.~Bernal,}
\author[a]{F.~Boudjema,}
\author[d]{S.~Chakraborti,}
\author[e]{A.~Goudelis,}
\author[f]{A.~Pukhov}
\affiliation[a]{LAPTh, CNRS, USMB, 9 Chemin de Bellevue,  F-74940 Annecy, France}
\affiliation[b]{School of Physics and Astronomy, University of Southampton\\
Southampton SO17 1BJ, United Kingdom}
\affiliation[c]{New York University Abu Dhabi\\
PO Box 129188, Saadiyat Island, Abu Dhabi, United Arab Emirates}
\affiliation[d]{IPPP, Department of Physics, Durham University, Durham, DH1 3LE,  United Kingdom}
\affiliation[e]{Laboratoire de Physique de Clermont Auvergne (UMR 6533), CNRS/IN2P3\\
Univ. Clermont, Auvergne, 4 Av. Blaise Pascal, F-63178 Aubière Cedex, France}
\affiliation[f]{Skobeltsyn Inst.\ of Nuclear Physics, Moscow State Univ., Moscow 119992, Russia} 

\emailAdd{belanger@lapth.cnrs.fr}
\emailAdd{a.belyaev@soton.ac.uk}
\emailAdd{nicolas.bernal@nyu.edu}
\emailAdd{boudjema@lapth.cnrs.fr}
\emailAdd{sreemanti.chakraborti@durham.ac.uk}
\emailAdd{andreas.goudelis@clermont.in2p3.fr}
\emailAdd{alexander.pukhov@gmail.com}

\abstract{We present \MO~7, a major upgrade of the \MO\ package for the computation of dark matter observables in generic models. This release introduces a generalized treatment of the Boltzmann equations, allowing for user-defined modifications of the Hubble expansion rate, entropy evolution, and non-thermal dark matter production from late-decaying cosmological components, thereby extending the framework beyond the standard radiation-dominated cosmology. The relic density can now be computed in scenarios such as low-temperature reheating, early matter domination, and kination. The new version also improves the treatment of sub-GeV dark matter, in particular annihilation into light mesons through scalar mediators, and provides updated spectra for indirect detection. Several experimental and observational constraints have been implemented or revised, including CMB bounds from Planck on energy injection during recombination and Fermi-LAT limits from dwarf spheroidal galaxies. For direct detection, a recast of recent LZ results has been included, and the code now takes into account  effective electromagnetic couplings of spin-$1/2$ and spin-1 dark matter. Collider observables have also been extended through the implementation of CMS dilepton resonance constraints on $Z'$ mediators. Additional improvements include a more flexible treatment of effective relativistic degrees of freedom and an updated LHAPDF interface.}

\begin{document}
\begin{flushright}
\end{flushright}
\maketitle

\newpage
\section*{PROGRAM SUMMARY}
\begin{description}
\item{{\it Program title:}} \MO~7
\item{{\it Developer's repository link:}} \url{https://micromegasdm.github.io/}
\item{{\it Licensing provisions:}} GNU General Public License 3 (GPL)
\item{{\it Programming language:}} C and Fortran
\item{{\it Journal reference of previous version:}} \href{https://doi.org/10.1016/j.cpc.2024.109133}{Comput. Phys. Comm.  299 (2024)109133}
\item{{\it Does the new version supersede the previous version?:}} Yes
\item{{\it Reasons for the new version:}} This new version of \MO\ allows to compute the dark matter relic density in non-standard cosmological scenarios. Moreover, it improves the treatment of sub-GeV dark matter and includes updated experimental and observational constraints on dark matter.
\item{{\it Summary of revisions:}} This version includes new routines to compute the abundance of dark matter in non-standard cosmological scenarios, such as low-temperature reheating, early matter domination, and kination. This version also improves the treatment of sub-GeV dark matter, in particular annihilation into light mesons through scalar mediators, and provides updated spectra for indirect detection. Moreover several experimental and observational constraints have been implemented or revised including CMB bounds from Planck on energy injection during recombination, Fermi-LAT limits from dwarf spheroidal galaxies, a recast of recent LZ limits and CMS dilepton resonance constraints on $Z'$ mediator. Finally, the code now takes into account  effective electromagnetic couplings of spin-$1/2$ and spin-1 dark matter for the computation of elastic scattering of dark matter on nucleons.
\item{{\it Nature of problem:}} The precise determination of the relic density of dark matter implies strong constraints on dark matter  models. The predictions for the dark matter abundance typically assume the standard radiation-dominated cosmology. We provide a public code  to perform a precise computation of the relic density for generic extensions of the standard model in non-standard cosmological scenarios.
\item{{\it Solution method:}}  We solve  Boltzmann equations that allow for  user-defined modifications of the Hubble expansion rate, entropy evolution, and non-thermal dark matter production from late-decaying cosmological components.

\item{{\it Typical running time:}}
On a laptop equipped with an Intel Core i5-6300U processor
(2.4~GHz, 2 physical cores and 4 hardware threads), the sample parameter
files distributed with micrOMEGAs require on average an execution time of  4.5~s. Some examples with freeze-in or
full-analysis  can take up to several minutes.
These timings do not include the one-time generation of matrix-element
libraries, which are subsequently stored on disk. A single micrOMEGAs
calculation is not parallelised over several CPU cores.

\item{{\it Performance Scaling:}}
For large parameter scans dominated by relic-density calculations,
we tested the scaling on IRIDIS up to about 200 cores and obtained around
98\% parallel efficiency. The scaling becomes significantly poorer when
indirect-detection calculations or external collider codes such as
HiggsBounds/HiggsSignals are included, mainly because of process generation,
external program calls and filesystem activity. The best scaling is therefore
obtained for relic-density calculations, with the more I/O-intensive stages
treated separately.

\end{description}

\section{Introduction}
The search for particle dark matter (DM) continues to motivate the development of computational tools capable of accurately predicting the relic density of DM, their various  scattering rates for direct detection, their indirect detection signals and cosmological signatures across a broad range of theoretical frameworks~\cite{Belanger:2001fz, Gondolo:2004sc, Belanger:2006is, Backovic:2013dpa, Ambrogi:2018jqj, Arbey:2018msw, Bringmann:2018lay, Palmiotto:2022rvw, Alguero:2023zol, Capucha:2025iml}. \MO\ has long provided a flexible platform for evaluating DM observables in generic extensions of the Standard Model (SM)~\cite{Belanger:2001fz, Belanger:2006is, Alguero:2023zol}. In this new release, we introduce a series of significant improvements that extend the applicability of the code to  non-standard cosmological histories  as well as to some light DM scenarios. Moreover, several constraints from astroparticle searches for DM, cosmic microwave background (CMB) measurements, direct detection experiments, and collider searches for new physics have been added or updated to incorporate the latest experimental results.

The major development of this release is a generalized treatment of the Boltzmann equations that govern the thermal evolution of DM. In the standard cosmological scenario with one DM component, the relic abundance is obtained by solving a single Boltzmann equation assuming radiation domination, entropy conservation, and a fixed relation between the temperature of the SM plasma and the Hubble expansion rate. This standard framework becomes  inadequate for scenarios motivated by inflationary reheating or hidden-sector dynamics, in which the early Universe may have followed a more complex thermal history.

The new implementation in \MO\ lifts these assumptions by allowing user-defined modifications to the expansion rate and entropy evolution. Concretely, the code now supports cosmologies in which the Hubble parameter receives additional contributions—such as from a period of early matter domination, or a kination era—as well as scenarios where entropy is injected through late decay or annihilation of heavy fields. For some early investigations, see Refs.~\cite{Giudice:2000ex, Fornengo:2002db, Pallis:2004yy, Pallis:2005hm, Gelmini:2006pw}. Such processes naturally arise in a large class of ultraviolet completions: supersymmetric models with late-decaying moduli~\cite{Acharya:2009zt, Kane:2015jia}, freeze-in and secluded dark sectors, theories with early kination~\cite{Zeldovich:1961sbr, Spokoiny:1993kt, Joyce:1996cp, Ferreira:1997hj, DEramo:2017gpl, Barman:2021ifu, Biondini:2023hek, Belanger:2025ack}, or models with additional relativistic degrees of freedom that temporarily dominate the energy density~\cite{Turner:1983he}.

By incorporating these effects directly into the Boltzmann equations, \MO\ now enables reliable and self-consistent relic-density calculations in settings where the standard freeze-out or freeze-in  picture is distorted or fundamentally altered. This includes cases where the annihilation rate becomes inefficient due to enhanced Hubble expansion or where the freeze-out temperature is shifted by entropy injection.  The resulting framework provides a flexible and robust tool to explore a wide range of early-Universe histories that are phenomenologically compelling yet were previously difficult to treat within a unified numerical code. The code also permits the user to update or replace the underlying tables for the effective relativistic degrees of freedom, $g_\text{eff}(T)$ and  $h_\text{eff}(T)$, allowing for a consistent treatment of cosmologies with additional light particles.

This new version of \MO\ also improves the modelling of light (sub-GeV) DM, a regime of increasing interest following both experimental progress and theoretical motivations. In particular, the code now includes a dedicated treatment of annihilation into light mesons for DM coupled through a scalar mediator, ensuring that hadronic effects below the GeV scale are properly accounted for. These developments are accompanied by an extended set of tables for DM-induced  positron yields down to masses below 2~GeV, relevant for both indirect detection signatures  and cosmological constraints  from the CMB. Moreover, new tables for photon, positron, and antiproton spectra have been added. The code also incorporates updated CMB constraints on light DM, using the standard approach for energy deposition efficiencies~\cite{Slatyer:2015jla}. In addition, \MO\ now provides indirect detection limits from Fermi-LAT observations of dwarf spheroidal galaxies~\cite{Calore:2018sdx, Alvarez:2020cmw}, which constitute an important probe of $s$-wave annihilation.

On the direct detection side, the package has been extended to  the  evaluation of the DM elastic scattering rate on nucleons with photon exchange. In particular, the effective electromagnetic dipole moments of DM for spin-1/2 DM and the dipole and quadrupole moments for spin-1 DM are now properly taken into account. We also include the most recent limits from the LZ experiment, implemented through a dedicated recast of the collaboration's latest results. 

Finally, collider-based constraints have  been extended. In particular, limits on $Z'$ mediators have been updated. Structural improvements include a new interface with recent versions of LHAPDF.

\begin{figure}[h!]
	\begin{centering}
	\includegraphics[width=1.\columnwidth]{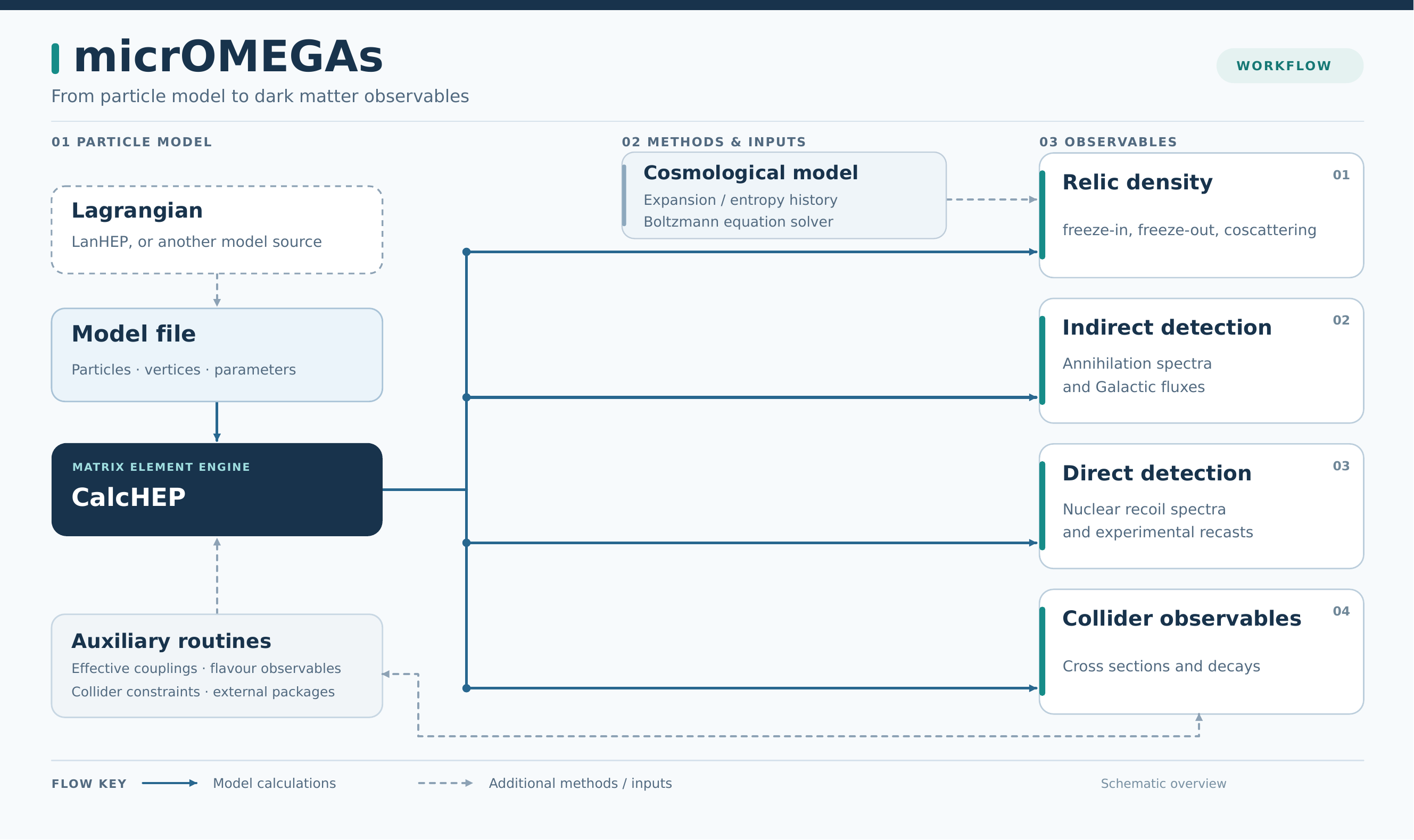}
    \end{centering}
	\caption{Flowchart of the \MO\ code.}
	\label{fig:flow}
\end{figure}

These new developments ensure consistent comparisons between cosmology, astrophysics, and collider probes across the viable parameter space. The code work flow is shown in Fig.~\ref{fig:flow}. In summary, the new features  of \MO~7 include:
\begin{itemize}
    \item{} A generalization of the Boltzmann equations to take into account non-standard cosmological scenarios. This major addition is represented by the component "cosmological model" in the \MO\ flowchart.
    \item{} Relic density for sub-GeV DM including annihilation into mesons in the case of a scalar mediator.
    \item{} CMB constraints on light DM.
    \item{} Possibility to update the tables for $g_\text{eff}$ and $h_\text{eff}$ to include new degrees of freedom.
    \item{} Updated collider limits on $Z'$ from CMS dilepton searches.
    \item{} Generalization of the direct detection routine to include  effective electromagnetic couplings of DM.
    \item{} Recasting of recent LZ results from direct DM searches.
    \item{} New tables for particle spectra relevant for indirect detection are added, thus four different sets of tables  are available to the user.
    \item{} The tables for DM-annihilation-induced $\gamma$, $e^+$, $\nu$ are extended to DM masses as low as 0.5~GeV when DM annihilation is into quark  pairs. For DM annihilation into meson pairs, spectra for photons and positrons are available for DM masses starting at the meson mass. These are relevant both for deriving cosmological and indirect detection constraints on light DM. 
    \item{} Indirect detection limits from Fermi-LAT observations of  Dwarf galaxies.
    \item{} Updated interface to LHAPDF.
\end{itemize}

This paper is organized as follows. In section~\ref{sec:lowTrh} non-standard cosmological scenarios  are discussed, along with the corresponding Boltzmann equations. Section~\ref{sec:lightDM} presents the calculation of the relic density for light DM interacting with the SM through a scalar mediator. Section~\ref{sec:DD} describes a new contribution to the elastic scattering rate that accounts for potential couplings of DM to photons. Section~\ref{sec:limits} concerns new or updated constraints on DM models from CMB anisotropies, indirect detection, and collider searches. All new and updated \MO\ routines are described in Section~\ref{sec:micro}. A sample output is given in Section~\ref{sec:output}, and Section~\ref{sec:conclusions} contains our conclusions. Finally, Appendix ~\ref{appendix} contains some practical applications of the new functionalities. 

\section{Dark matter relic density in non-standard cosmological histories} \label{sec:lowTrh}
In the very first release of \MO~\cite{Belanger:2001fz}, the description of the thermal evolution of the Universe was naturally based on a minimalist picture involving (at least) two underlying assumptions: \textit{i}) at the onset of the radiation domination (RD) era, the SM plasma was characterized by a very high temperature, much higher than  all relevant mass scales in the underlying microscopic theory and \textit{ii}) radiation domination continued uninterrupted with (radiation) density $\rR(T) \propto T^4$, and hence a Hubble expansion rate $H(T) \propto T^2$, with $T$ being the radiation temperature. In such a context, as long as the interactions between DM and the visible sector are strong enough, the former is produced thermally. This feature has an important practical consequence: given a particle physics model for DM (in the first release this was the minimal supersymmetric model, MSSM), the computation of  the DM relic density does not require knowledge of the initial conditions of the Universe; namely, the final outcome is independent both of the initial temperature of the plasma (the ``reheating temperature'' $\Trh$) and of the initial DM abundance. This is due to the fact that the DM particles $\chi$ (which are stable and electrically neutral) attain {\it thermal equilibrium} with the SM plasma, hence erasing all memory of the initial conditions. When the annihilation rate of DM into SM particles drops below the expansion rate of the Universe $H$, DM decouples and \textit{freezes out} while being non relativistic at a freeze-out temperature $T_{\rm fo} \sim M_{\chi}/25-M_{\chi}/20$, where $M_{\chi}$ is the mass of the DM candidate. One then also arrives at the important result that in the standard freeze-out scenario the relic density is inversely proportional to the annihilation rate of DM into SM particles. This picture gave rise to the WIMP paradigm, in which the DM particles have interactions of the same order as the SM weak couplings and masses around the weak scale.

A first break with these assumptions was introduced in \MO~5.0~\cite{Belanger:2018ccd}, in which DM particles with extremely small couplings, FIMPs, were considered. Although these particles do not attain thermal equilibrium, the correct relic density can be achieved through an alternative production mechanism, the so-called \textit{freeze-in} mechanism. Contrary to conventional freeze-out, in freeze-in the initial conditions \textit{can} matter: first, the initial dark matter abundance must be specified. Secondly, if the underlying DM-SM interactions are non-renormalizable or if $\Trh$ becomes comparable or smaller than one of the mass scales in the theory, the predicted DM abundance depends on its precise value. However, in this version of \MO\ it was still assumed that the SM entropy, $S$, is conserved.

This property fails already in the case where some field $\phi$, which filled the Universe with an energy density $\rp$, decays. Besides the fact that this new component makes the expansion rate larger than in the standard scenario at a fixed temperature, its late decay, characterized by a lifetime $\Gp^{-1}$, injects entropy into the SM bath. The width, $\Gp$,  specifies  a reheating temperature $\Trh \propto \sqrt{m_P\, \Gp}$, which can be quite low,  contrary to the tacit assumption in the standard scenario where  this temperature is very high, in particular, for freeze-out scenarios well above the freeze-out temperature. In fact, this reheat temperature can  be lower than the freeze-out temperature, and  even much lower, as long as it does not disturb BBN (Big Bang Nucleosynthesis), and one has to ensure that $\Trh$ remains above a few MeV, as required by BBN~\cite{Sarkar:1995dd, Kawasaki:2000en, Hannestad:2004px, Barbieri:2025moq}.

An interesting example of low-reheating temperature scenarios is the case where $\phi$ is the inflaton field~\cite{Bassett:2005xm, Allahverdi:2010xz, Amin:2014eta, Lozanov:2019jxc, Lozanov:2020zmy, Barman:2025lvk} which we  generalize to  any late decaying field. The field $\phi$ is often taken as having the properties of non relativistic, pressure-less, $p_\phi=0$, matter. Here we will consider the more general case of a fluid with an effective equation of state during reheating $p_\phi=w\, \rp$ relating the pressure $p_\phi$ and its energy density $\rp$.

This late-decaying field can also decay, either directly or through cascade decays, into  a  certain number of DM particles. In this case, in addition to thermal production, this new non-thermal contribution further modifies the DM abundance, its subsequent evolution, and ultimately the resulting relic density of DM.

During reheating, the injection of entropy by the decaying field  changes the standard relation between  $T$, the temperature that tracks the SM bath, and $a$ the Friedmann-Lemaître-Robertson-Walker cosmic scale factor. We allow for the possibility that the interaction rate, $\Gp$, need not be constant but depends on the scale factor. This is achieved by adding an extra parameter $\alpha$ which, during  the epoch of $\phi$  domination, results in the power-law dependence of the bath temperature
\begin{equation}
    T(a) \propto \frac{1}{a^\alpha}\,.
    \label{T(a)}
\end{equation}
In the standard scenario where the SM entropy, $S = s\, a^3$, is conserved (constant), and in the radiation domination era, this amounts to the familiar scaling law with $\alpha = 1$.\footnote{Barring factors from the effective degrees of freedom contributing to the entropy.}

\subsection{The background}
The evolution of the background in the presence of the field $\phi$ described above is governed by the set of coupled Boltzmann equations for the energy density $\rp$  and the entropy density $s$ of the SM thermal bath~\cite{Giudice:2000ex, Fornengo:2002db, Gelmini:2006pw, Allahverdi:2020bys, Batell:2024dsi} 
\begin{align}
    &\frac{d\rp}{dt} + 3\, (1 + w)\, H\, \rp = - \Gp\, \rp\,, \label{eq:rho}\\
    &\frac{ds}{dt} + 3\, H\, s =  \frac{\Gp\, \rp}{T}\,, \label{eq:s}
\end{align}
where $H$ is the Hubble expansion rate of the Universe, which now takes into account the SM radiation and the field $\phi$,
\begin{equation}\label{eq:H}
    H^2 = H_\phi^2 + H_\text{SM}^2 = \frac{8\pi\, \rp}{3\, m_P^2} + \frac{8\pi\, \rho_\text{SM}}{3\, m_P^2}\,,
\end{equation}
with $m_P \simeq 1.2 \times 10^{19}$~GeV the Planck mass and ignoring the (subdominant) contribution of the DM energy density. The SM energy density is given by
\begin{equation}
    \rho_\text{SM}(T) = \rR(T) = \frac{\pi^2}{30}\, g_\text{eff}(T)\, T^4
    \label{rhosm}
\end{equation}
while the entropy $s$ 
\begin{equation}
    s(T) = \frac{2\pi^2}{45}\, h_\text{eff}(T)\, T^3,
    \label{eq:heff}
\end{equation}
where $\gs$ and $\gss$ are the number of degrees of freedom that contribute to $\rho_\text{SM}$ and $s$, respectively. Interestingly, Eq.~\eqref{eq:heff} serves as a definition of the temperature.

The interaction rate $\Gp(a)$ is generically not constant but can be a function of the scale factor; we parametrise it as~\cite{Bernal:2025fdr}
\begin{equation} \label{eq:Gamma1}
    \Gp(a) = \Gp(a_I) \left(\frac{a_I}{a}\right)^\frac{8\alpha - 3(w+1)}{2}= \Gp(a_I) \left(\frac{a_I}{a}\right)^{-\gamma}
\end{equation}
to produce a SM bath with a temperature $T$ that scales during reheating as in Eq.~\eqref{T(a)}, where we have denoted by $a_I$ the value of the scale factor at the onset of reheating. Note that the models in which $\Gp(a)$ is taken to be constant correspond to the choice 
\begin{equation}
    \gamma = \frac{3 (w+1)-8 \alpha}{2} = 0\,.
    \label{eq:csteGp}
\end{equation}
A notable example is the early-matter-dominated scenario with $w = 0$ and $\alpha = 3/8$~\cite{Giudice:2000ex}. Besides, in the limit $\Gp \to 0$, Eq.~\eqref{eq:s} corresponds to entropy conservation. In the same limit  $\rp(a)$ scales as $ a^{-3(1+w)}$, and drops faster than the radiation density $\rR(a) \propto a^{-4}$, for $w > 1/3$.

It is interesting to note that a viable reheating, that is, an eventual onset of the SM radiation domination, requires that at some point $\rR(a) > \rp(a)$, which in turn implies
\begin{equation} \label{eq:noRH}
    \alpha \leq \frac34\, (1 + w)\,.
\end{equation}

To solve the system of equations~\eqref{eq:rho} and~\eqref{eq:s}, in \MO\ we find it convenient to express them  in terms of the scale factor $a$ and the comoving variables $z_\phi \equiv \rp \times a^{3(1+w)}$ and $z_s \equiv s^{4/3} \times a^4$ as
\begin{align}
    \frac{dz_\phi}{da} &= - \frac{\Gp}{H\, a}\, z_\phi\,,  \label{eq:zphi}\\
    \frac{dz_s}{da} &=  \frac43\, \frac{\Gp}{H}\, \frac{s^{1/3}}{T}\, \frac{z_\phi}{a^{3w}}\,. \label{eq:zs}
\end{align}
These equations are solved numerically using the routine \verb|getInflDecay| or \verb|getInflDecayPlus|, described in Section~\ref{sec:functions}. These functions tabulate and save in memory the dependence of $\rp$ and $s$ on the scale factor $a$, which can then be used in the calculation of the relic density. In \MO\ the reheating temperature $\Trh$ is defined as the temperature at which the SM radiation density and the density of the field $\phi$ become equal, 
\begin{equation}
    \rho_{\phi}(\Trh) =\rho_\text{SM}(\Trh)\,,
    \label{eq:defTrh}
\end{equation}
and its value is found numerically.

Without loss of generality, in \MO\ the input parameters that define the system of equations for the evolution of the background  and the required boundary conditions are the following free parameters 
\begin{equation}
    {\tt H_{SM}}\equiv H_\text{SM}(a_I),\, {\tt H_\phi} \equiv H_\phi(a_I),\, \Gp=\Gp(a_I),\, \alpha,\, \text{and}\, w \quad (\text{with}\; a_I=1)
    \label{inputsMO}
\end{equation}
where the first three parameters must be given in units of GeV. These parameters are the arguments of the function \verb|getInflDecayPlus|. A function \verb|getInflDecay|, with fewer arguments, is also provided for models with a restricted number of parameters (see Section~\ref{sec:functions} later). The set of $5$ parameters in Eq.~\eqref{inputsMO} to describe a non-standard, generalized background allows for a wide range of applications. It  can, for example, be implemented  for  a large class of inflaton models~\cite{Bernal:2024yhu, Bernal:2024jim, Bernal:2024ndy, Bernal:2025fdr, Banik:2025olw, Bernal:2025fcl}. The inflaton inspired scenario most widely studied involves a heavy inflaton oscillating at the minimum of a quadratic potential, whose energy density redshifts as non-relativistic matter ($w = 0$): an early matter dominated era. In this regime, the inflaton undergoes perturbative decays into pairs of SM particles, leading to $\alpha = 3/8$~\cite{Giudice:2000ex}. This reheating phase naturally emerges in well-motivated inflationary models, including Starobinsky inflation~\cite{Starobinsky:1980te} and a variety of polynomial inflation scenarios~\cite{Drees:2021wgd, Bernal:2021qrl, Drees:2022aea, Bernal:2024ykj}. These fall into the category of models that satisfy Eq.~\eqref{eq:csteGp} for which the decay width $\Gp$ is constant.\\   
In the context of $\alpha$-attractor inflation models~\cite{Kallosh:2013hoa, Kallosh:2013maa}, the inflaton can oscillate about the minimum of a monomial potential $V(\phi)\propto \phi^n$ with $n\geq 2$. In such cases, the effective equation-of-state parameter during reheating is $w = (n -2)/(n + 2)$~\cite{Turner:1983he}. The value of $\alpha$ also depends sensitively on the mechanism by which the inflaton transfers its energy to radiation, particularly on the structure of the inflaton–matter couplings~\cite{Co:2020xaf, Garcia:2020wiy, Xu:2023lxw, Barman:2024mqo}. For example, perturbative inflaton decay into bosons yields $\alpha = 3/(2(n+2))$, while decay into fermions gives $\alpha = 3(n-1)/(2(n+2))$ for $n < 7$ and $\alpha = 1$ for $n > 7$. If reheating proceeds through the inflaton annihilation into bosons through contact interactions, we obtain $\alpha = 9/(2(n+2))$ for $n \geq 3$. In scenarios where annihilation is mediated by a light scalar in the $s$-channel, resonant effects can lead to $\alpha = 3(7-2n)/(2(n+2))$ for bosonic final states and $\alpha = 3(5-n)/(2(n+2))$ for fermionic final states~\cite{Barman:2024mqo}. If the mediator is, instead, heavy, annihilation into fermions gives $\alpha = 1$~\cite{Barman:2024mqo}. Additionally, certain models allow for a nearly constant temperature during reheating, corresponding to $\alpha = 0$~\cite{Co:2020xaf, Barman:2022tzk, Chowdhury:2023jft, Cosme:2024ndc}.\\
Finally, if the energy density of the field $\phi$ redshifts faster than that of free radiation~\cite{DEramo:2017gpl} (that is, if $w > 1/3$) its depletion does not require decay ($\Gp=0$),  or annihilation. Here, $\alpha = 1$ as in kination-like ($H(a) \propto a^{-3}$) set-ups~\cite{Zeldovich:1961sbr, Spokoiny:1993kt, Joyce:1996cp, Ferreira:1997hj} which is also the same scaling of the temperature as in the standard scenario.

\subsection{Dark matter}
Specifying the model for DM and its interactions, in the cosmological background defined in the previous section, the DM yield can be computed by accounting for contributions from both the DM plasma and the possible decay of the field $\phi$ into a certain number, $k$, of DM particles, $\chi$. The evolution of the DM number density $n_\chi$ is governed by
\begin{equation} \label{Boltzmann}
    \frac{dn_\chi}{dt} + 3\, H\, n_\chi = -\langle v\sigma_A \rangle \left( n_\chi^2 - n_\text{eq}^2 \right) - \frac{1}{2}\langle v \sigma_S \rangle \left( n_\chi^2-n_\chi\, n_\text{eq} \right) + b\, \Gp\, \rp\,,
\end{equation}
where $n_\text{eq}(T)$ is the DM number density at equilibrium and the additional parameter
\begin{equation}\label{eq:b}
    b = \frac{1}{m_\phi}\, \sum_k k\, \text{Br}(\phi\to k\, \chi  )\,,
\end{equation}
where $m_\phi$ is the mass of the field $\phi$, which is assumed to remain constant during reheating. Note that we take into account  both the cross-sections for DM annihilation,  $\sigma_A$, and  semi-annihilation, $\sigma_S$. 

We re-express Eq.~\eqref{Boltzmann} through the variables $z_\chi\equiv n_\chi\, a^3$ and $\bar z_\chi\equiv n_\text{eq}\, a^3$,
\begin{equation} \label{eq:dzda}
    \frac{dz_\chi}{da} =-\frac{1}{H a^4} \left[\langle v\sigma_A\rangle \left(z_\chi^2-\bar{z}_\chi^2  \right) +\frac{1}{2}\langle v\sigma_S\rangle \left(z_\chi^2-z_\chi \bar{z}_\chi  \right) - b\, \Gp\, z_\phi\, a^{3(1-w)}\right].
\end{equation}
This equation is solved numerically by the routine \verb|darkOmegaInfl|, described in Section~\ref{sec:functions}.

Following the same approach, we have also  generalized the solution of the Boltzmann equations for $N$-component DM to include the new cosmological background. For each DM component ($\chi_\mu$, $\mu=1, \dots, N$), the evolution equations described in Ref.~\cite{Alguero:2023zol} are extended to include the  decay term
\begin{equation}
    \frac{d z_{\chi_\mu}}{da} = -\frac{1}{H a^4} \left[ \sum_{\alpha\le\beta, \gamma\le\delta} C_{\alpha\beta}\, z_{\chi_\alpha}\, z_{\chi_\beta} \langle v\sigma_{\alpha\beta\gamma\delta}\rangle (\delta_{\mu\alpha} + \delta_{\mu\beta} - \delta_{\mu\gamma} - \delta_{\mu\delta}) - b_\mu\, \Gp\, z_\phi\, a^{3(1-w)} \right],
\end{equation}
where $C_{\alpha\beta} \equiv 1 - \frac12\, \delta_{\alpha\beta}$, $\langle v\sigma_{\alpha\beta\gamma\delta}\rangle$ is related to the total number of events per unit space time volume for processes involving a pair of DM particles into SM bath particles that have the index $\delta$, $\gamma = 0$. The relation between the direct and inverse annihilation processes reads
\begin{equation}
   C_{\alpha\beta}\, \langle v\sigma_{\alpha\beta\gamma\delta}\rangle\, \bar{z}_{\chi_\alpha}\, \bar{z}_{\chi_\beta} =  C_{\gamma\delta}\, \langle v\sigma_{\gamma\delta\alpha\beta}\rangle\, \bar{z}_{\chi_\delta}\, \bar{z}_{\chi_\gamma}\,,
\end{equation}
and in particular for  $ \chi_\alpha\chi_\beta\to $ SM,SM we get 
\begin{equation}
    C_{\alpha\beta}\, \langle v\sigma_{\alpha\beta00}\rangle\, \bar{z}_{\chi_\alpha}\, \bar{z}_{\chi_\beta}  = \frac12\, \langle v\sigma_{00 \alpha\beta}\rangle\, z_0\, z_0\,.
\end{equation}
Here we follow  the notation in Ref.~\cite{Alguero:2023zol} which contains more details on the solution of $N$-component equations. Finally, the branching ratio of the scalar into each DM component is defined as in Eq.~\eqref{eq:b}, $b_\mu \equiv \sum_k k\, \text{Br}(\phi\to k\, \chi_\mu)/m_\phi$. 

\subsection{Validation of the code}
\begin{figure}[b!]
	\centering
	\includegraphics[width=0.49\columnwidth]{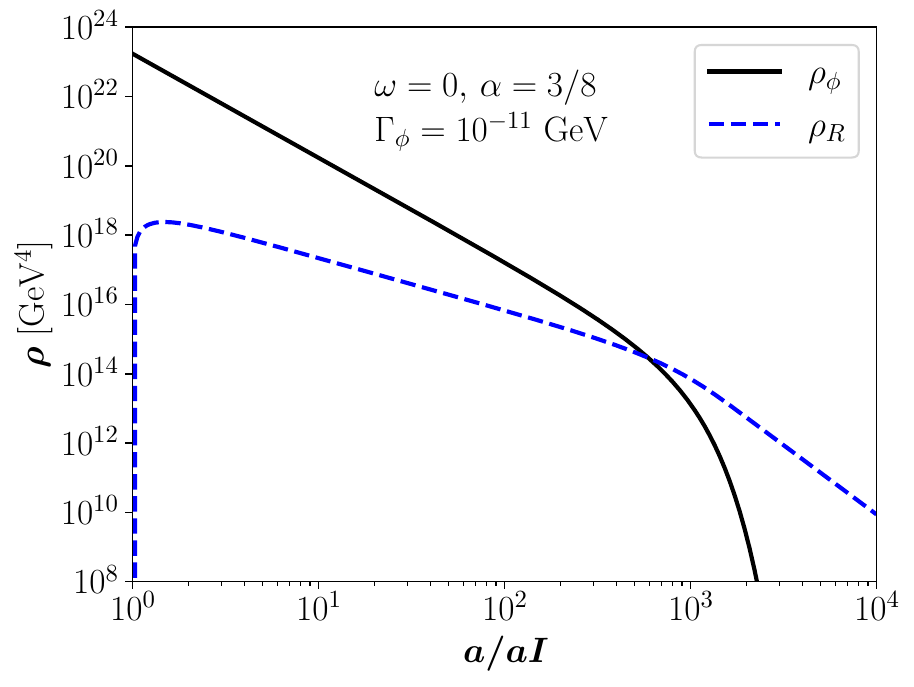}
	\includegraphics[width=0.49\columnwidth]{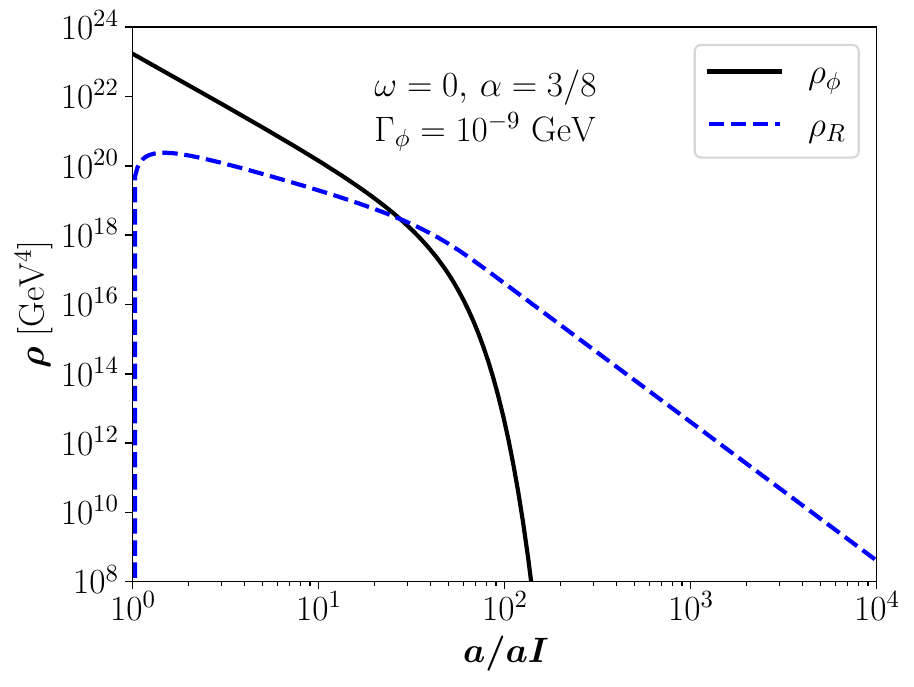}
	\includegraphics[width=0.49\columnwidth]{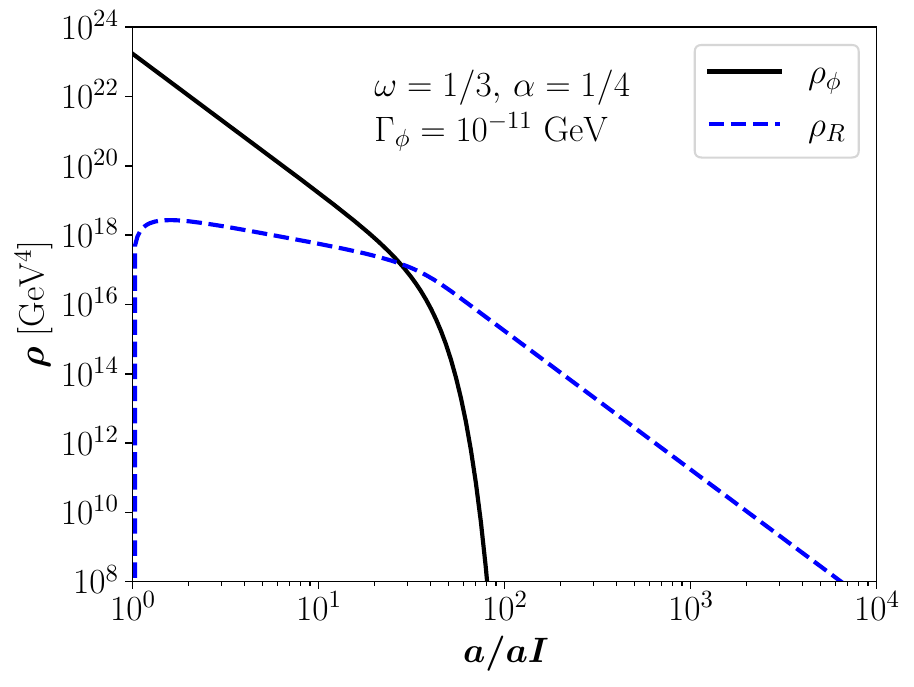}
	\includegraphics[width=0.49\columnwidth]{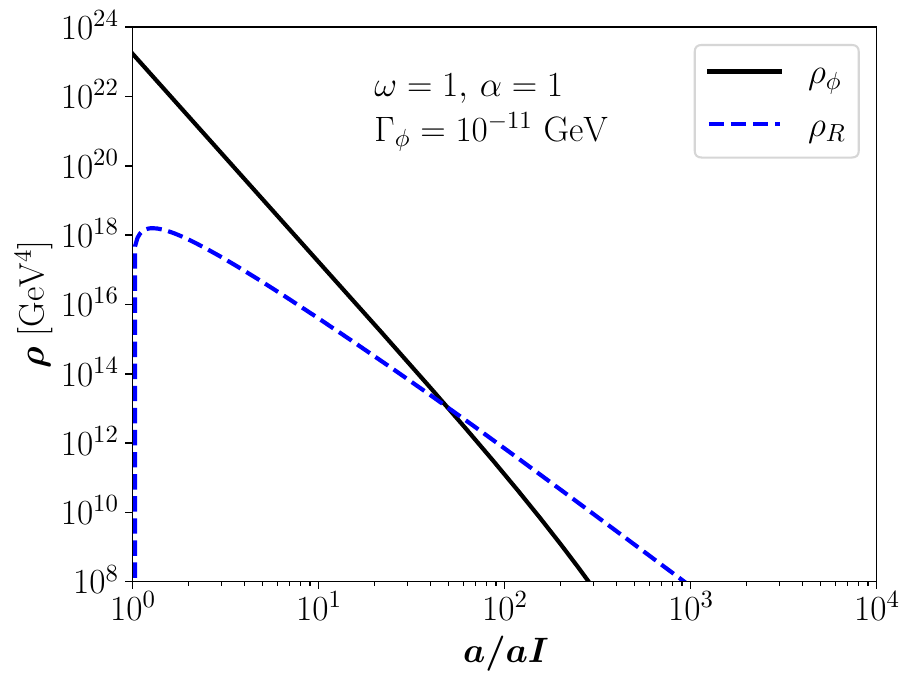}
	\caption{Evolution of the energy densities, for $w = 0$ and $\alpha = 3/8$ ($\gamma=0$, top), $w = 1/3$ and $\alpha = 1/4$ ($\gamma=1$, bottom left), and $w = \alpha = 1$ $(\gamma=-1$, bottom right), for $H_\phi = 10^{-7}$~GeV and $H_\text{SM} = 0$~GeV.}
	\label{fig:cosmo}
\end{figure}
To validate the code, we compared the evolution of $\rp$ and $\rR$ obtained with \MO\ (using the function \verb|getInflDecayPlus| described in Section~\ref{sec:functions}) with independent numerical solutions presented in Refs.~\cite{Belanger:2024yoj, Belanger:2025ack, Belanger:2026ctm}. This comparison was performed for different choices of $\alpha$ and $w$, and we found excellent agreement.

Figure~\ref{fig:cosmo} displays representative examples of the evolution of the SM energy density (dashed blue) and the  energy density of $\phi$ (solid black), calculated with \MO. The upper panels correspond to an early matter-dominated era ($w=0$ and $\alpha=3/8$, falling into the category of constant, scale independent, $\Gp$, see Eq.~\eqref{eq:csteGp}), with $\Gp = 10^{-11}$~GeV (left) and $\Gp = 10^{-9}$~GeV (right). As expected, increasing $\Gp$ shortens the lifetime of $\phi$, causing $\rR$ and $\rp$ to intersect at a smaller value of the scale factor $a$, and consequently leading to a higher reheating temperature $\Trh$. Recall that in the radiation dominated era $\rR \gg \rp$, for larger values of $a/a_I$, $T(a) \propto 1/a$. Increasing $\Gp$ also means that the depletion of $\phi$ due to its decay into SM particles occurs earlier and is much faster (exponential decrease) when the decay is effective, $\Gp \sim H$ around $\Trh$. The exponential decrease exacerbates the one caused by dilution. This also explains why the radiation density increases quite dramatically, more so as $\Gp$ is larger.

In addition, we have also explored other cosmological scenarios such as a radiation dominated era with $w=1/3$ and $\alpha=1/4$ (bottom left) and kination $w = \alpha = 1$ (bottom right), where in both cases the width is scale-dependent and further affects the evolution of $\rR$ and $\rp$. Before its effective decay, $\rp(a)$ scales as $a^{-3(1+w)}$ (dilution term). This explains part of the ever sharper decrease of $\rp$ with increasing values of $w$ in Fig.~\ref{fig:cosmo}.

\begin{figure}[htb!]
	\centering
	\includegraphics[width=0.49\columnwidth]{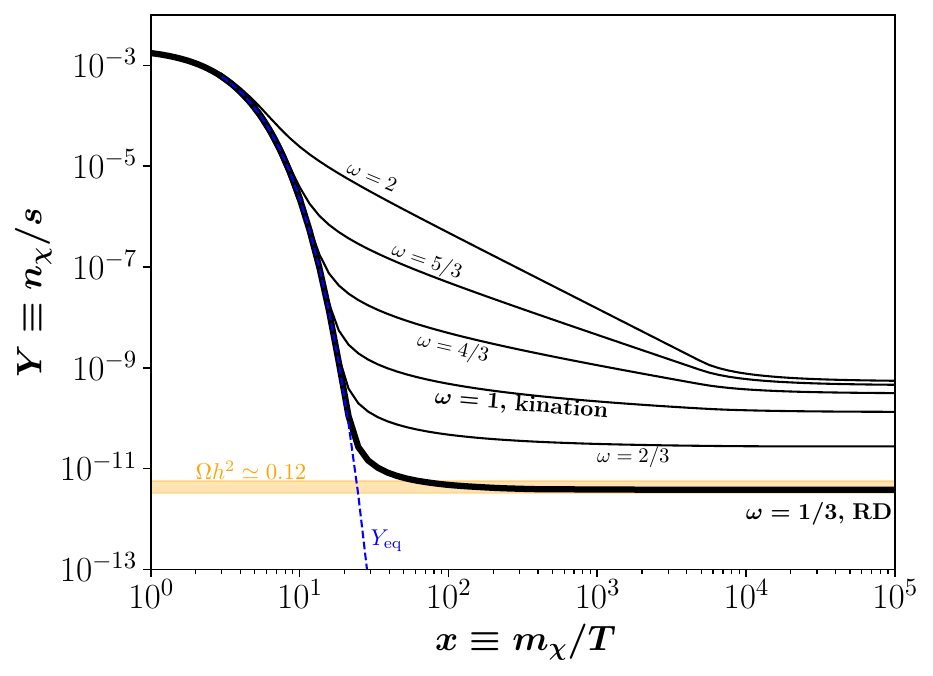}
	\caption{Evolution of the abundance for $m_\chi=100$~GeV, $\langle\sigma v\rangle = 1.7 \times 10^{-9}$~GeV$^{-2}$, $\Gp = 0$, $\alpha=1$ for different choices of $w$.}
	\label{fig:relentless}
\end{figure}

Incorporating DM and the prediction of \MO\ for the relic density, we have compared  the DM relic density by taking a fixed constant annihilation cross-section for DM in the case of  the relentless cosmological scenario of Ref.~\cite{DEramo:2017gpl}. Again we found excellent agreement, see Fig.~\ref{fig:relentless}. Here, for each choice of $w$ the values of $H_\text{SM}$ and $H_\phi$ are tuned such that $\Trh=20$~MeV. For example, for $w=1$, $H_\text{SM} = 10^{-9}$~GeV and $H_\phi = 4\times 10^{-3}$~GeV lead to $\Trh = 20$~MeV and $\Tmax=26$~TeV. In this example the value of the cross-section is tuned so that the value of the relic density in the radiation dominated cosmology is compatible with  PLANCK observations while other cosmological scenarios with $w>1/3$ lead to overabundant DM. In contrast, cosmologies with $w>1/3$ offer the possibility to regain compatibility with observation for models where DM is predicted to be underabundant in the standard cosmological scenario~\cite{Belanger:2025ack}.

Finally, we used two simple models, the inert doublet (IDM) and $Z'$ portal (ZpPortal) models to illustrate the impact of different choices of cosmological parameters  on the abundance of DM. In the first model, the  DM is a WIMP, while we choose the parameter of the second model to get a FIMP DM. Full details can be found in the Appendix.

\section{The case of sub-GeV dark matter annihilating into hadrons} \label{sec:lightDM}
We have implemented in \MO\ a dedicated framework for computing the thermal relic abundance with DM masses below the GeV scale and interacting with the SM through a light scalar mediator which couples to mesons. In this kinematic regime, DM annihilation proceeds predominantly into light pseudoscalar mesons, and therefore requires a treatment of non-perturbative QCD effects beyond the standard quark-level description used at higher DM masses.

The new facility extends \MO' range of applications by providing a consistent and flexible description of the scalar–meson interactions relevant for sub-GeV phenomenology in the case of a scalar that mixes with the SM Higgs. The effective coupling of the scalar mediator to the pseudoscalar mesons $\pi$, $K$, $\eta$ is implemented through three independent parametrisations, designed to capture the range of theoretical uncertainties associated with hadronic matrix elements in the non-perturbative regime. The first relies on leading-order chiral perturbation theory ($\chi$PT-LO), where analytical expressions for the scalar couplings to mesons  follow directly from the explicit chiral symmetry–breaking terms in the QCD Lagrangian.  This parametrisation is expected to be reliable for scalar masses below 500 MeV. The second employs energy dependent  scalar form factors constructed using dispersion relations constrained by unitarity and experimental input~\cite{Winkler:2018qyg}. This parametrisation incorporates  the contributions of low-lying scalar resonances, providing a more accurate description in the vicinity of hadronic thresholds. The form factors entering the light scalar decay width  are extracted from Fig.~2 in Ref.~\cite{Winkler:2018qyg}. The third provides an alternative dispersive construction, based on a different set of form factor inputs~\cite{Blackstone:2024ouf}. Here, two different sets of form factors, which depend on the low-energy matching procedure, are obtained using the code \verb|hipsofcobra|~\cite{cobra}. This code also provides an estimate of their uncertainties that can reach an order of magnitude for the width of the scalar~\cite{Blackstone:2024ouf}.

These parametrisations are made available through  user-selectable flags in the input file.  An example of the implementation of scalar couplings to mesons in the case of a simplified model with fermionic or scalar DM coupling to a Higgs portal mediator can be found in the directory \verb|LightScalar| of the public \MO\ distribution.  A more detailed description is provided separately in Ref.~\cite{Belanger2026}. 

\section{Direct detection} \label{sec:DD}
\subsection{Recasting of LZ}
Direct-detection limits are typically presented under a set of simplifying assumptions: a single DM component, contact interactions mediated by heavy particles, and a standard halo velocity distribution. However, many well-motivated models violate one or more of these assumptions. In theories with multi-component DM, the local density of each species—and thus its scattering rate—differs from the canonical value assumed in experimental analyses. Models with light mediators modify the momentum dependence of the scattering amplitude, altering both the recoil spectrum and the relative sensitivity of different experiments. Likewise, non-standard velocity distributions, arising for example from baryonic feedback, dark substructure, or late-time interactions, can significantly change the predicted event rate. For these reasons, a faithful comparison between theoretical predictions and experimental data requires recasting the published limits using the correct density fraction, interaction structure, and velocity distribution appropriate to the model under consideration.

\begin{figure}[htb]
    \centering
    \includegraphics[width=0.8\textwidth]{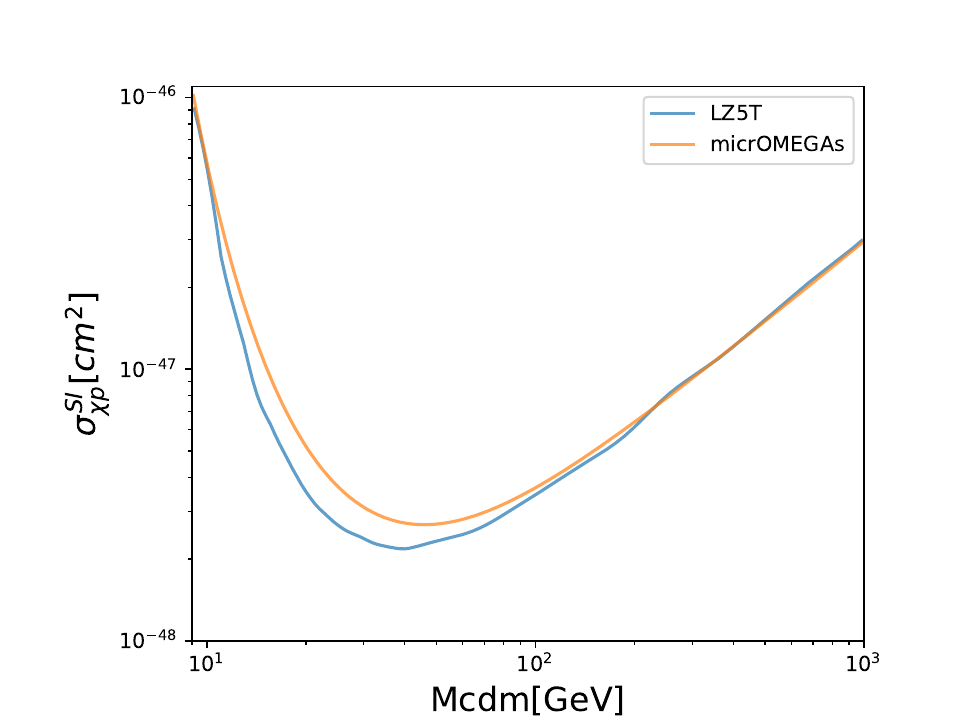}
    \caption{Comparison of the  limit on spin-independent elastic scattering cross-section for DM on protons obtained by LZ5T (blue) with the one from  \MO\ recasting (orange).}
    \label{fig:LZ}
\end{figure}

The strongest limits from SI direct detection in the ${\cal O}(100)$~GeV mass range are obtained by the LZ collaboration. The number of events is given by
\begin{equation}
    N_\text{ev}= L \int{f} \frac{dN}{dE_R} dE_R
\end{equation}
where $L$ is the exposure, $L = 5500$~kg$\times 60$~days in Ref.~\cite{LZ:2022lsv} and $L = 3300$~kg$\times 365$~days in Ref.~\cite{LZ:2024zvo}. The efficiency $f$ is given by the experimental collaboration. In \MO\ we implemented the functions $f$ provided in Ref.~\cite{LZ:2022lsv} (green curve in Fig.~2) and Ref.~\cite{LZ:2024zvo} (yellow curve in Fig.~2) corresponding to single-scatter reconstruction and analysis cut. For the data of 2022, Ref.~\cite{LZ:2022lsv}, we introduce a cut at $E_R>17$~keV to remove the background at high recoil energies. For this conservative recast, we assume that there are no background events and no signal events, so that the 90\% confidence level (C.L.) limit corresponds to $N_\text{ev} = 2.3$. For electroweak-scale DM with a standard velocity distribution,  the recast limits match very well those of LZ5T from the 2022 and 2024 analyses, except for masses below 20 GeV where the maximum difference is close to 50\%, see Fig.~\ref{fig:LZ}. However, note that our limit is conservative in the sense that it is always less restrictive than the LZ one.

\subsection{Photon exchange}
In the present version of \MO\ we have modified the function that computes the elastic scattering rate of DM on nuclei in order to take into account the case where DM interacts with the photon through higher-dimensional electromagnetic operators. The implementation currently supports both dipole (magnetic or electric) for spin-1/2 DM as well as  quadrupole interaction for spin-1. For example, DM interactions with charged particles may induce an interaction with photons at the loop level. The Lagrangian with higher order operators must be included in the model file by the user. 

We have improved the treatment of the DM-nucleus interaction for the case of a photon exchange between the DM and the nucleus in the function \verb|nucleusRecoil| that computes the recoil energy distribution for DM scattering on a nucleus. This function is used by \verb|DD_pval| to determine the exclusion in a given model, as described in the previous subsection for  the exclusion by the latest LZ results. In previous versions~\cite{Belanger:2020gnr}, only the case of a millicharged Coulomb interaction was treated, while the more realistic case with dipole  or quadrupole interactions caused by loop diagrams was not included. The new \verb|nucleusRecoil| function takes into account the $\chi\bar\chi\gamma$ vertex implemented in the \MO\ model file. The function includes all diagrams with massive particle exchange, photon exchange ($\chi\bar\chi\gamma$), and their interference.

The photon exchange diagram is not included in the  function \verb|nucleonAmplitudes| that computes the amplitudes for elastic scattering on the nucleons. This function now returns 4 if a $\chi\bar\chi\gamma$ vertex is detected in the model file. This should be considered as a warning that the DM scattering on the nucleus should be used to compare with the experimental limits. The {\tt dipoleDM} directory contains an example of the treatment of millicharge DM or dipole moments for direct detection limits.

Note that in addition to the photon, other light mediators are treated in the function \verb|nucleonAmplitudes|. Since this function always assumes that the mediator mass is much larger  than the transferred momentum and that this approximation fails for the mediator mass $\lesssim 100$~MeV, the routine returns 1(2) if a light mediator is found for SI (SD) interactions. This indicates to the user that DM scattering on the nucleus where the dependence on the transferred momentum is taken into account should be used instead.  

This addition enables \MO\ to compute direct-detection signals while keeping the relevant momentum-transfer dependence and thus can be applied to  a broad class of dark-sector models with electromagnetic moment interactions.

\section{Improved limits on dark matter and new particles} \label{sec:limits}
\subsection{CMB-ION}
Particles produced from  DM annihilation at a redshift around 1000 would inevitably affect the process of recombination, leaving an imprint on Cosmic Microwave Background (CMB) anisotropies and polarization~\cite{Slatyer:2015jla}. 

This indirect DM constraint from CMB observations is abbreviated as CMB-ION. Its sensitivity is comparable and complementary to conventional indirect (ID) DM searches searching for signals from DM annihilation products at present times. One of the main advantages of the CMB-ION constraint (which includes both signal and background considerations) is that it does not depend on propagation of DM annihilation products through the galactic medium. Thus, the CMB-ION constraint is  much more robust than the ID constraint discussed in the next section. Note, however, that the CMB-ION constraint assumes that DM annihilation is  $s$-wave and not suppressed at low velocities; it should be applied with care whenever the annihilation rate at recombination differs from the low-velocity $s$-wave value, for example, in $p$-wave annihilations, resonance-enhancements, threshold dependencies, or in cases with Sommerfeld-enhancement.

\begin{figure}[htb]
    \centering
    \begin{subfigure}[b]{0.47\textwidth}
        \includegraphics[width=\textwidth]{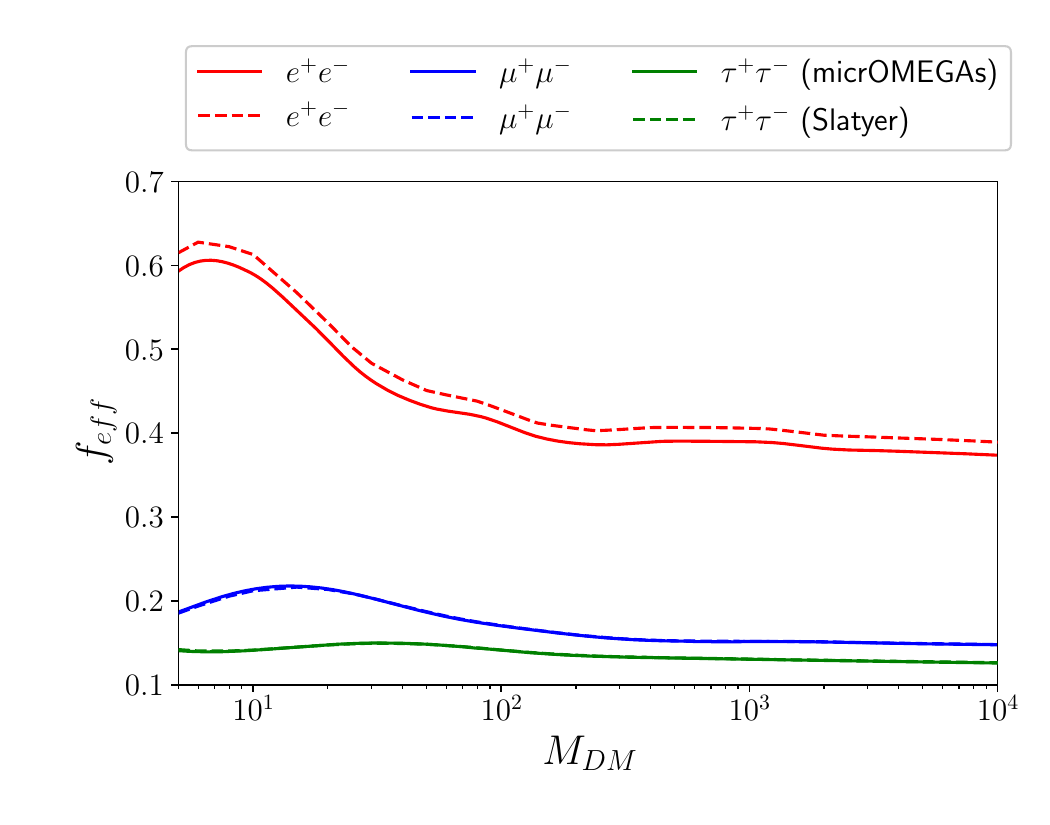}%
        \caption{}
    \end{subfigure}%
    \begin{subfigure}[b]{0.47\textwidth}
        \includegraphics[width=\textwidth]{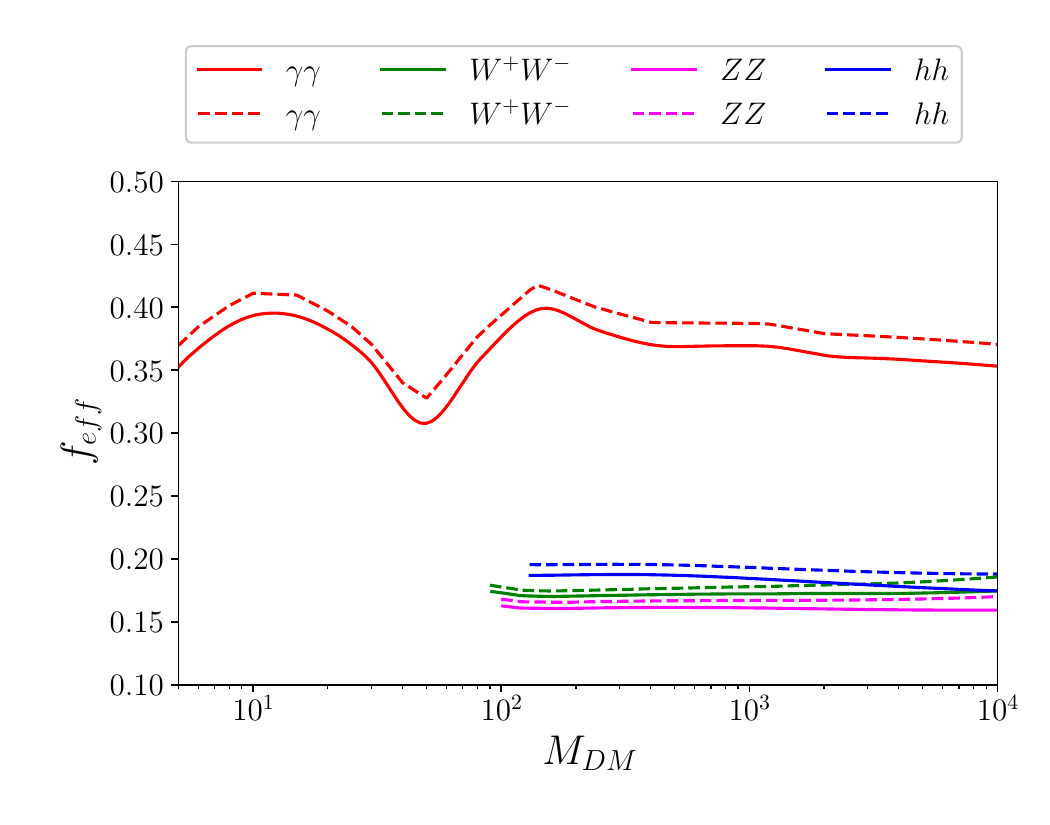}%
        \caption{}
    \end{subfigure}\\
    \begin{subfigure}[b]{0.47\textwidth}
        \includegraphics[width=\textwidth]{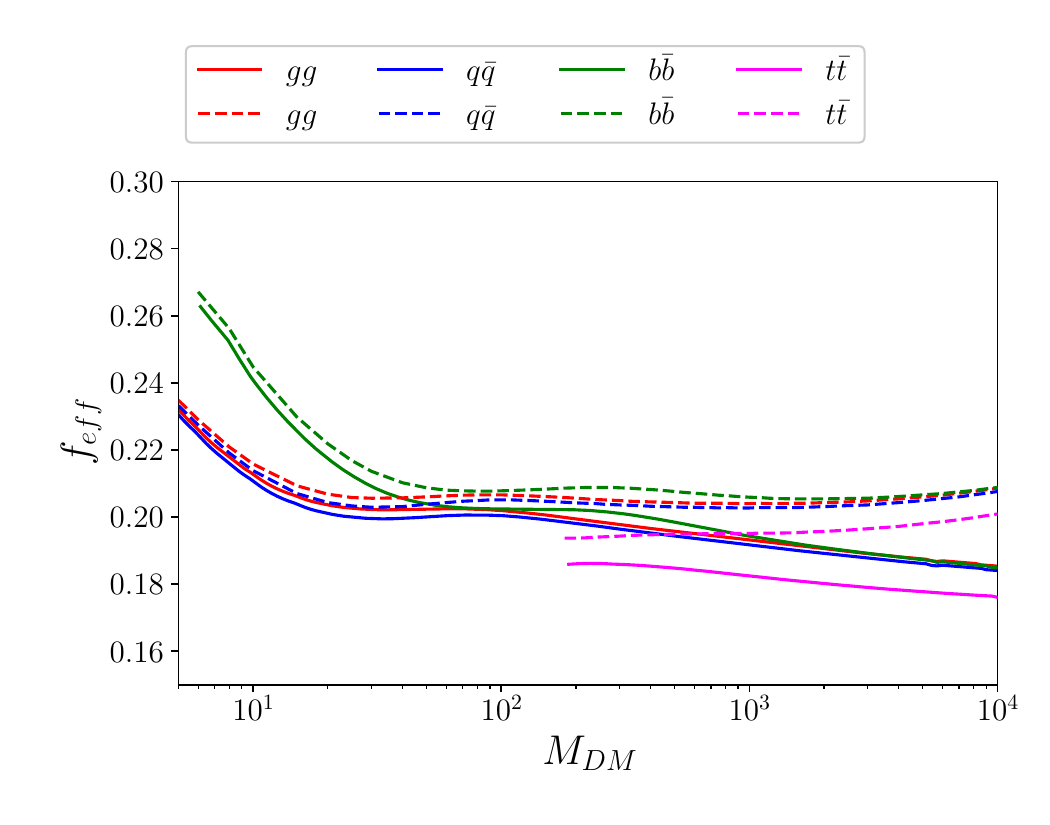}%
        \caption{}
    \end{subfigure}%
    \caption{\label{fig:feef}A comparison of $f^i_\mathrm{eff}(m_{\rm DM})$ from \MO\ (full lines)  and Ref.~\cite{Slatyer:2015jla} (dashed lines) for various DM annihilation channels. In both cases the {\tt PPPC4DMID} spectra are used~\cite{Ciafaloni:2010ti, Cirelli:2010xx}.}
\end{figure}

The effective parameter constrained by the CMB anisotropies is
\begin{equation}
    p_{\mathrm{ann}} =  \sum_i f_\mathrm{eff}^i \times \text{Br}_i \times \frac{\langle\sigma v \rangle}{m_{\mathrm{DM}}},
    \label{eq:pann}
\end{equation}
where  $m_{\mathrm{DM}}$ is the mass of the DM particle, $\langle\sigma v\rangle$ its thermally averaged annihilation cross-section at the CMB time, $f^i_\mathrm{eff}$ is the fraction of the energy deposited from particles of type $i$ (from DM annihilation) into the intergalactic medium (IGM) around the redshifts to which the CMB anisotropy data are most sensitive, ($z \sim 600$), while $\text{Br}_i$ is the branching ratio for DM annihilation in each particular channel. Annihilation channels into non-SM particles  are also included. For these the positron and photon energy spectra are obtained by computing the branching fractions into SM particles and boosting the resulting spectra. Hence, we effectively include annihilation into 4 SM particles final states keeping only on-shell channels~\cite{Belanger:2010gh}. The best limit on $p_{\mathrm{ann}}$ is given in the latest Planck collaboration analysis~\cite{Planck:2018vyg} as
\begin{equation}
    p_{\mathrm{ann}}<3.2\times 10^{-28} {\rm cm}^3 {\rm s}^{-1} {\rm GeV}^{-1}    \quad (95 \%, \text{Planck}\,\,\mathrm{TT},\mathrm{TE},\mathrm{EE+lowE+lensing+BAO}).
    \label{eq:pann-limit}
\end{equation}
In \MO, we evaluate $f^i_\mathrm{eff}$ using the following integral
\begin{equation} 
    f^i_\mathrm{eff}(m_{\mathrm{DM}}) = \frac{1}{2 m_{\mathrm{DM}}}\, \int_0^{m_{\mathrm{DM}}} E \mathrm{d}E \left[ 2 f^{e^+ e^-}_\mathrm{eff}(E)  \left(\frac{dN}{dE} \right)_{e^+}^i + f^{\gamma}_\mathrm{eff}(E)  \left(\frac{dN}{dE} \right)_{\gamma}^i \right], 
    \label{eq:feef}
\end{equation} 
where $E$ is the energy of the positron or photon, while $f^{e^+ e^-}_\mathrm{eff}(E)$ and  $f^{\gamma}_\mathrm{eff}(E)$ are the tabulated functions from Ref.~\cite{Slatyer:2015jla, Slatyer:2015kla} for the fractions of electron and photon energies deposited in IGM as a function of their initial energy. In Eq.~\eqref{eq:feef}, the quantities $\left(\frac{dN}{dE}\right)^i_{e^+}$ and $\left(\frac{dN}{dE}\right)^i_\gamma$ denote the energy spectra of positrons and photons, respectively, for a particular DM annihilation channel $i$. As a cross check,  in Fig.~\ref{fig:feef} we present a comparison between $f^i_\mathrm{eff}(m_{\rm DM})$ as computed in \MO\ and the results obtained in Ref.~\cite{Slatyer:2015jla} for various DM annihilation channels, assuming the same $\left(\frac{dN}{dE}\right)^i_{e^+,\gamma}$ spectra taken from {\tt PPPC4DMID}~\cite{Ciafaloni:2010ti, Cirelli:2010xx}. 
The agreement is within a few percent for the leptonic and bosonic channels. For the quark and gluon final states, larger differences appear at high $m_{\rm DM}$, as shown in Fig.~\ref{fig:feef}(c).\footnote{In discussion with T.~Slatyer, this discrepancy was traced to a bug in the original code used to produce the corresponding curves of Ref.~\cite{Slatyer:2015jla}; the corrected behaviour is obtained in \MO.}
Small variations on the constraint can be expected depending on the tables for the DM annihilation-induced positron and photon spectra chosen by the user.

\subsection{Indirect detection}
The present version of \MO\ further introduces several important improvements to its indirect detection-related functionalities. 
It incorporates the  Fermi-LAT limits from dwarf-spheroidal galaxies, allowing direct comparison between predicted gamma-ray fluxes and observational constraints. For this we rely on  the code~\cite{calore_2021_5592836} which is based on a data-driven approach to estimate background and derive limits from Fermi-LAT based on  observations of 25 dwarf galaxies~\cite{Calore:2018sdx, Alvarez:2020cmw}. A discussion on the background modelling and the J-factor dependence can be found in Ref.~\cite{calore_2021_5592836}. Because the total photon spectra coming from all DM pair-annihilation channels is computed by \MO\ (with {\tt calcSpectrum}) this feature is particularly useful for models where DM annihilates into several SM final states, or in BSM particles. For the latter, the photon spectrum is constructed automatically by considering the decay chain of BSM particles into SM final states. Moreover, this function allows for an easy comparison of the theoretical predictions with Fermi-LAT limits for multi-component DM~\cite{Belanger:2021lwd}.

In this version, the tables describing the energy spectra of positrons from DM annihilations into all pairs of SM particles have been extended to cover DM masses below 2 GeV.  This enables reliable predictions for light DM and is essential to implement  both CMB and indirect detection constraints in this mass range. Moreover, \MO\ now provides four independent sets of particle-spectra tables relevant for indirect searches; in particular, the new tables from {\tt CosmiXs}~\cite{Arina:2023eic} have been added. Lastly, the hadronic uncertainties on the spectra obtained by \textsc{Pythia}~8~\cite{Bierlich:2022pfr} are accessible. Together, these new features give the users greater flexibility and improved accuracy when estimating gamma-ray and charged-particle yields across a wide range of DM masses. 

\subsection[Collider limits on $Z'$]{\boldmath Collider limits on $Z'$}
The limits from LHC experiments on searches for new particles are mostly obtained through an interface with {SModels} which includes several searches for new physics with missing transverse energy (MET) as well as monojet searches~\cite{Altakach:2024jwk}.  In addition, \MO\ contains a routine that provides a limit on a new vector boson, $Z'$, which can act as a mediator between DM and SM particles in certain models. 

We have implemented the 95\% C.L. limit on a new $Z'$ boson as obtained in the CMS analysis of dilepton resonance production based on 140~fb$^{-1}$ of Run~2 data~\cite{CMS:2021ctt}. Internally, \MO ~uses \texttt{CalcHEP} to extract the vector and axial couplings of the $Z'$ to up- and down-type quarks and the leptonic branching ratio $\mathrm{Br}(Z'\!\to \ell^+\ell^-)$; these quantities are then used to construct the effective parameters $c_u$ and $c_d$ employed in the CMS analysis. The procedure follows the formalism of Ref.~\cite{Accomando:2010fz}, where the 95\%~C.L. limits on dilepton resonances are expressed in terms of these two coefficients, which encode the effective strength of the $Z'$ couplings to quark initial states in dilepton production.

For the Lagrangian
\begin{equation}
    \mathcal{L} = Z'_\mu \bar{q}\gamma^\mu(g_v + g_a\gamma^5)\,q ,
\end{equation}
$c_q = 2\bigl(g_V^{q\,2} + g_A^{q\,2}\bigr)\,\mathrm{Br}(Z'\!\to \ell^+\ell^-)$, where $\ell = e,\mu$. The CMS analysis assumes universal $Z'$ couplings to charged leptons, implying $\mathrm{Br}(Z'\!\to e^+e^-) = \mathrm{Br}(Z'\!\to \mu^+\mu^-) = \mathrm{Br}(Z'\!\to \ell^+\ell^-)$, and the experimental limit is derived from the combined $e^+e^-+\mu^+\mu^-$ channel. In this formalism, $c_q$ ($q=u,d$) is the effective coefficient that determines the partonic production strength of the $Z'$ through initial-state quarks of flavour $q$ in the CMS dilepton analysis; it depends only on the $Z'$ couplings to quarks and on its leptonic branching ratio and does not involve any cross-section calculation.

The couplings $g_V^q$, $g_A^q$ and the leptonic branching ratio entering $c_q$ are obtained using the tools presented in Sections~18.5 and~18.6 of the manual provided with the code in \verb|man/manual_7.1.pdf|. In the CMS analysis, the exclusion contour for a fixed $M_{Z'}$ corresponds to a straight line in the $(c_d,c_u)$ plane. We find the boundary points of this line $[c^{\rm Max}_d(M_{Z'}),0]$ and $[0,c^{\rm Max}_u(M_{Z'})]$. Then \verb|ZpLimCMS| returns
\begin{equation}
    c_d/c_d^\text{Max}(M_{Z'}) \;+\; c_u/c_u^\text{Max}(M_{Z'}) .
\end{equation}

Each model point is mapped onto a unique position $(c_u,c_d)$ in this plane. Comparison with the CMS straight-line boundary yields the value returned by {\tt ZpLimCMS}: values larger than~1 signal exclusion at 95\%~C.L.\ or stronger, while values below~1 correspond to allowed points. This provides a fast and robust test of generic $Z'$ scenarios against the CMS dilepton limits without requiring explicit cross-section calculation, detector simulation, or event generation, provided that the $Z'$ couplings are correctly specified in the \texttt{CalcHEP} model files.

\section{Description of \texttt{\bf micrOMEGAs}' functions} \label{sec:micro}
\subsection{Non-standard cosmologies} \label{sec:functions}
New functions are available in \MO\ to compute the DM relic density after solving the dynamics of the background. All the routines listed below contain \verb|Infl| in their name, since they were originally designed for the case of inflaton decay; however, they apply to  a generic component $\phi$ as discussed in Section~\ref{sec:lowTrh}:\\

\noindent $\bullet$ \verb|getInflDecayPlus(Hsm, HI, Gamma, alpha, w, &Trh, &Tmax, &aEnd)|\\
is the function that solves Eqs.~\eqref{eq:zphi} and~\eqref{eq:zs} in the generalized case. The first 5 arguments of the function correspond to the 5 input parameters defined in Eq.~\eqref{inputsMO}. The code returns the temperature when the energy density of the inflaton becomes equal to the energy density of the SM particles, \verb|Trh|, the maximum temperature that can be reached during the evolution, \verb|Tmax|,  and \verb|aEnd|  the value of the scale factor $a$ at a temperature \verb|Tend|. This temperature is defined by the user as a global parameter.\\

\noindent $\bullet$ \verb|getInflDecay(HI, Gamma, &Trh, &Tmax, &aEnd)|\\
This function solves Eqs.~\eqref{eq:zphi} and~\eqref{eq:zs} in the particular cases where $\alpha=3/8$, $w =0$ and $H_{\text{SM}}=0$. The other arguments have the same meaning as in the previous function. \\

Both functions \verb|getInflDecayPlus| and \verb|getInflDecay| return 0 when a solution to the respective differential equations is found, or 128 otherwise. For both functions, the  evolution of the Universe is stored in the  tabulated  functions \verb|Ta(a)|, \verb|Ha(a)|, \verb|rhoSMa(a)|, \verb|rhoIa(a)| which represent respectively the temperature of the SM bath, the total expansion rate, the energy  density of the SM particles, and the energy density of the field $\phi$. These functions are then used to calculate the relic density of DM. The argument of these functions is in  the  interval $a \in [1, \text{aEnd}]$. The final value of the  scale factor, \verb|aEnd|, is defined automatically by the final temperature of the SM bath, \verb|Tend|.\footnote{ If the condition $\rho_R>\rho_\phi$ is never satisfied, the code returns $\Trh$ = NaN, there is no error or warning issued.}\\

\noindent $\bullet$ \verb|aT(T)|\\
returns the scale factor $a$ at a given temperature $T<$~{\tt Tmax}. {\tt aT(T)} ={\rm NaN} when $T$ exceeds {\tt Tmax}. In general, for $T<$~{\tt Tmax}  there are two values of $a$ that correspond to a given temperature. {\tt aT} returns the largest  of the two  which corresponds to the cooling time  of the Universe.\\
  
To solve for the relic density, we have two functions:\\
\noindent $\bullet$ \verb|darkOmegaInfl(b,Beps, &isWIMP, &err) |\\
solves the DM evolution Eq.~\eqref{Boltzmann} where {\tt b} is the branching fraction for the decay of the  inflaton into $N$ DM particles divided by  the inflaton mass in GeV, see Eq.~\eqref{eq:b}. This function  uses \verb|Ta|, \verb|rhoIa|, and \verb|Ha| obtained by \verb|getInflDecay| or \verb|getInflDecayPlus|. The main return value of this routine is the DM relic density $\Omega_\chi h^2$. The auxiliary output parameters are {\tt isWIMP} ({\it integer}) and {\tt err} ({\it integer}). The parameter {\tt isWIMP= 1} for a WIMP scenario, that is if $\Omega$ decreases when the annihilation cross section increases. Otherwise {\tt isWIMP=0}. The error code {\tt err} indicates whether there is a problem with the solution of the differential Eqs.~\eqref{eq:zphi} and~\eqref{eq:zs}, in which case the 8$^\text{th}$ bit of the error code {\tt err} is 1, or with the solution of Eq.~\eqref{eq:dzda}, then the 9$^\text{th}$ bit is 1. The lowest bits contain information about problems with numerical integration and should be treated as warnings. The parameter \verb|Beps| has the same meaning as in other \verb|darkOmega| routines; it determines the condition to include Boltzmann suppressed channels~\cite{Belanger:2001fz}. The DM abundance is stored in an array and is available through the functions {\tt Za(a)}, \verb|Ya(a)| while the equilibrium abundance is stored in {\tt ZaEq(a)}, \verb|YaEq(a)|.\\

\noindent $\bullet$ \verb|darkOmegaNInfl(b, Beps,  &err) |\\
solves the DM evolution Eq.~\eqref{eq:dzda} for $N$-component DM~\cite{Alguero:2023zol}. This function has the same argument and output as above. The DM abundance is stored via the functions {\tt ZaN(a,ch)}, \verb|YaN(a,ch)|, while the equilibrium abundance is stored in {\tt ZaNEq(a,ch)}, \verb|YaNEq(a,ch)| where \verb|ch| is a text which specifies the channel, for example {\tt ch="2"}. {\tt b} represents an array of coefficients that describe the branching fraction of the inflaton decay into DM components, see Eq.~\eqref{eq:b}. In particular, the first coefficient (that is, b[0]) corresponds  to the branching fraction of $\phi$ into the first DM component CDM[1].

The user has the possibility to define the parameters that enter \verb|getInflDecayPlus| and \verb|darkOmegaInfl| in the input  file. It can be done using the following records after adding the  numerical values, 
\begin{verbatim}
    #! HsmInfl ... 
    #! HInfl ... 
    #! GammaInfl ...
    #! wInfl ...
    #! alphaInfl ..
    #! BrInfl ... 
\end{verbatim}
where {\tt HsmInfl}, {HInfl}, {\tt GammaInfl}, {\tt alphaInfl}  and \verb|wInfl| correspond respectively to $H_{SM}$, $H_\phi$, $\Gamma$,  $\alpha,w$ in Eq.~\eqref{inputsMO} and \verb|BrInfl| is b in Eq.~\eqref{eq:b}. An example is given in  {\tt IDM/infl2.par}.

Since \MO\ can now treat three different  scenarios  for the calculation of the  relic density: {\it freeze-out}, {\it freeze-in} and modified cosmology denoted as  {\it inflaton}, the scenario can be selected automatically by specifying a flag in the input file :
\begin{verbatim}
#! FreezeOut
#! FreezeIn 
#! Inflaton 
\end{verbatim}
The first flag is applied  by default and can be omitted. The selection can also be made by the user in the main.c file.

\subsection{Contribution of BSM particles to entropy and energy density }
The effective entropy and energy density enter the Boltzmann equations for the DM abundance (see Eqs.~\eqref{rhosm} and~\eqref{eq:heff}); new functions allow the  user to include the contribution of new particles beyond the SM.\\

\noindent$\bullet$~\verb|h1eff(M/T,mu/T, eta)| \\
\noindent$\bullet$~\verb|g1eff(M/T,mu/T, eta)| \\
return the contribution of a single internal degree of freedom of a particle with mass \texttt{M} and chemical potential \texttt{mu} to the effective entropy (\texttt{hEff}) and energy (\texttt{gEff}) density functions. The argument \texttt{eta} specifies the particle statistics: \texttt{eta = 1} for bosons and \texttt{eta = -1} for fermions. To obtain the total contribution of a new particle species, the value returned by \verb|h1eff| or \verb|g1eff| must be multiplied by its internal degree of freedom $g_i$. By default, \MO\ includes only Standard Model particles in \texttt{hEff} and \texttt{gEff}; these functions allow the user to add the contribution of additional BSM particles present in the thermal bath to the total effective degrees of freedom entering the computation of the Hubble rate and entropy density. To do this, the user can create a new table in the format of the existing ones in the \verb|Data/hgeff| directory and then use \verb|loadHeffGeff("filename")|. The chemical potential is necessary  to treat models with lepton asymmetry, otherwise the user can set $\mu/T=0$.
 
The functions \\
\noindent$\bullet$ \verb|p1eff(M/T,mu/T,eta)|\\
\noindent$\bullet$ \verb|n1eff(M/T,mu/T,eta)|\\
can be used to calculate the  pressure and number density of one degree of freedom
\begin{eqnarray}
    p&=& \frac{\pi^2}{90}  {\rm p1eff}(M/T,\mu/T,\eta)T^4,\\
    n&=&  {\rm n1eff}(M/T,\mu/T,\eta)T^3.
\end{eqnarray}
These functions are normalized so that 
\begin{eqnarray}
    {\rm h1eff(0,0,1)}&=&{\rm g1eff}(0,0,1)={\rm p1eff}(0,0,1)=1,\\
    {\rm  n1eff}(0,0,1)&=&\frac{\zeta(3)}{\pi^2}.
\end{eqnarray}  

In the tables {\tt LM.thg} and {\tt DHS.thg} for degrees of freedom that are currently provided in \MO\ in the directory \verb|Data/hgeff|, the condition for entropy conservation,
\begin{equation}
    \frac{d\rho}{dT} = T\frac{d s}{dT}  
\end{equation}
is not always exact in the region of quark deconfinement. An improved solution that imposes entropy conservation is provided  in the files {\tt LMi.thg} and {\tt DHSi.thg}. The correction can be found in  the code {\tt mdlIndep/improveGeff.c } which outputs a comparison between the two sets of tables. The difference is only 0.5\%. 

\subsection{Constraints}
\subsubsection{Planck constraint on CMB anisotropies}
$\bullet$ \verb| PlanckCMB(vSigma, SpA, SpE) |\\
determines whether the DM model under consideration is consistent with Planck measurements of the CMB anisotropies caused by the injection of electrons and photons coming from DM annihilation~\cite{Slatyer:2015jla}. The spectra of DM annihilation into photons and positrons are provided in the arrays {\tt SpA} and {\tt SpE}.
 It is assumed that DM accounts for the full observed relic density, $\Omega h^2=0.12$. For multi-component DM, the fraction of each DM component described by the {\tt fracCDM} array is taken into account.
The input parameter {\tt vSigma} is assumed to represent the low-velocity $s$-wave annihilation cross section in cm$^3$/s units that remains effectively constant over the redshift range relevant for the CMB constraint. The routine should therefore not be applied directly when the annihilation rate has a significant velocity or redshift dependence. A return value greater than 1 indicates that the model is excluded at 95\% C.L. Note that if the DM model gives $\Omega h^2<0.12$ and one assumes that DM forms only a fraction {$\xi=\Omega h^2/0.12$ of the total DM in the Universe, then the value returned by \verb|PlanckCMB| should be scaled by $\xi^2$.

In the standard use of the routine, {\tt vSigma} is obtained from {\tt calcSpectrum}, which computes the annihilation cross section at vanishing relative velocity; this is appropriate for velocity-independent $s$-wave annihilation. If a model of elementary particles is not loaded and the user simply defines the spectra assuming specific annihilation channels, the DM mass entering the normalization is taken from the {\tt Mcdm} parameter. This allows for a model-independent test of the function {\tt PlanckCMB}.

In the file {\tt mdlIndep/CMB.c} we reproduce the upper row of Fig.~3 and the upper panel of Fig.~4  of Ref.~\cite{Slatyer:2015jla}, using the {\tt PPPC4DMID} spectra.

\subsubsection{Indirect detection}
$\bullet$  \verb|SpectraFlag| \\
is a switch to choose the tables that contain the spectra  for $\gamma,e^+,\bar{p},\nu$ from DM pair annihilation into two particle final states. This Flag has to be set before calling the functions {\tt calcSpectrum} and {\tt basicSpectra}.\\
{\tt SpectraFlag=0}  corresponds to new tables obtained with \textsc{Pythia}~8~\cite{Belanger:2010gh} for DM mass in the range  $2~{\rm  GeV} - 5~{\rm TeV}$ for $e^+,\bar{p},\nu$. For the photon and positron channels the tables  have  been extended to cover the range $110~{\rm MeV} - 5~{\rm TeV}$. These tables also include the spectra for polarized $W$'s and $Z$'s. To obtain the spectra generated by the transverse and longitudinal $W$'s substitute \verb|pdgN|$=24+'T'$ and $24+'L'$ correspondingly. In the same manner \verb|pdgN|$=23+'T'$ and $23+'L'$ provide the spectra produced by a polarized $Z$ boson.\\
{\tt SpectraFlag=1} corresponds to the spectra generated by \textsc{Pythia}~8~\cite{Amoroso:2018qga, Jueid:2022qjg} for DM mass in the range  $5~{\rm  GeV} - 5~{\rm TeV}$, here polarization is not taken into account.\\
{\tt SpectraFlag=2} corresponds to the spectra generated with {\tt PPPC4DMID}~\cite{Ciafaloni:2010ti, Cirelli:2010xx} for DM mass larger than 5~GeV. The spectra take into account polarized $W$ and $Z$,  as well as polarized leptons, \verb|pdgN|$= 11 +'R'$ or \verb|pdgN|$= 11 +'L'$,  the latter are however not supported in the current version of \MO.\\
Finally, {\tt SpectraFlag=3} corresponds to the  spectra  provided in  {\tt CosmiXs} for DM in the range  $5~{\rm  GeV} - 100~{\rm TeV}$~\cite{Arina:2023eic}. These spectra are generated with \textsc{Pythia}~8 and include electroweak corrections. These  spectra also included longitudinal/transverse $W$ and $Z$  as well as  left/right polarized  charged leptons.

The \textsc{Pythia}~8, {\tt PPPC4DMID}, and {\tt CosmiXs} tables contain spectra that start at  $x \gtrsim 10^{-9}$, while the default spectra {\tt SpectraFlag=0} start at  $x \gtrsim 10^{-7}$. To download  the  spectra including smaller energies, one has to recompile \MO\ after setting the parameter \verb| NZ|=295 in the  file {\tt include/micromegas.h}. 

The maximum values of the $x$ parameter in \textsc{Pythia}~8 and {\tt CosmiXs} tables are  0.94 and 0.9, respectively. For this reason,  the  spectral line for DM, DM$\to \gamma,\gamma$  which corresponds to a contribution  $2 \delta(1-x)$ in the photon spectra is not included, only the sub-dominant part of the photon spectra from photon radiation is included. In \MO\  the  $\delta$-like contribution to photon spectra in the $\gamma,\gamma$ channel is added to the \textsc{Pythia}~8 and {\tt CosmiXs} tables. In the same manner, we add a $\delta(1-x) $ contribution to the neutrino spectra for the $\nu,\bar{\nu}$ channel. In {\tt PPPC4DMID} the corresponding $\delta$-like contributions are simulated by  a linear function in the  interval $x\in[0.89,1]$. 

The QCD uncertainties that can impact the  annihilation spectra  were evaluated in Ref.~\cite{Jueid:2022qjg,Jueid:2023vrb}. For each DM mass, annihilation channel and choice of final state, Ref.~\cite{Jueid:2022qjg} have estimated three types of QCD uncertainties and provide tabulated spectra including these uncertainties. These tables are now incorporated in the code. We have defined three parameters that determine which uncertainties are taken into account,
\begin{itemize}
   \item{DHad:} uncertainties on the flux from the variations of the hadronisation model parameters.
   \item{DScale:} shower uncertainties on the flux from the variation of the shower evolution variable by a factor of 2.
   \item{DcNS:}  uncertainties  from the variations of the non-singular terms of the DGLAP splitting kernels. 
\end{itemize}
By default, these parameters are set to 1 and all types of uncertainties are taken into account; however, any one can be switched off by setting  the corresponding parameter to zero. The total uncertainties $\Delta^2(E)_{min,max}$ are obtained by taking the sum of the squares of the individual uncertainties. The global parameter {\tt spectUncert} determines whether or not uncertainties are included in the spectra. When {\tt spectUncert=0}, the default spectra without uncertainties, $(dN/dE)_{wo}$, are used, {\tt spectUncert=1} means that {\tt basicSpectra}  returns the maximal  spectra corresponding to $(dN/dE)_{wo}+\Delta(E)_{max}$  while { \tt spectUncert=-1} the minimal spectra corresponding to $(dN/dE)_{wo}-\Delta(E)_{min}$  are used. Note that although these uncertainties were calculated for the \textsc{Pythia}~8 spectra, \MO\ uses the same uncertainties for other spectra. One can access the spectra corresponding to the specified value of {\tt spectUncert} by calling the following function\\
$\bullet$ \verb|spectraUncertainty(Mass,pdgN,outN,Spectr)|\\
which has the same parameters as {\tt basicSpectra}. 

The test code  {\tt mdlIndep/basicSpectra.c} contains an example of a call to {\tt basicSpectra} and compares the four different spectra that are included in \MO, checks energy conservation, and shows uncertainties.

\noindent
$\bullet$ \verb|DwarfSignal(vSigma, SpA)|\\
determines whether the photon spectra {\tt SpA}  is compatible with the 95\% limits obtained from observations of dwarf spheroidal galaxies Carina, Fornax, LeoI, SegueI,  Draco, LeoII, Sculptor, Sextans~\cite{Bonnivard:2015xpq, Fermi-LAT:2016uux, Alvarez:2020cmw}. Here, {\tt vSigma} is the  DM annihilation cross section at low velocities ($\langle v\sigma\rangle$ in units of cm$^3$/s) as obtained by \verb|calcSpectrum|.  A return value, $r\ge 1$,  indicates  that the given model is excluded at 95\% confidence level.  The quantity $\langle v\sigma\rangle/r$ gives an estimate of the 95\% excluded cross section. However, note that since $0.01<r<100$, such an estimate fails if the initial $\langle v\sigma\rangle$ is more than a factor of 100 larger or smaller than the 95\% excluded value.

For a model independent test of this function, the user has to specify the value of the {\tt Mcdm} parameter before calling the function. For  example,  see the file {\tt mdlIndep/Dwarf.c}  which reproduces  the  {\it combined} curve in Ref.~\cite{Alvarez:2020cmw}, Fig.~5 (right).

\noindent
$\bullet$ \verb|calcSpectrumPlus(proc22, outP, Spect, &err)|  \\
calculates DM annihilation in 3-body and 4-body final states via the exchange of virtual particles. These processes are not taken into account by {\tt calcSpectrum}. Here {\tt proc22} specifies the   $2\to2$ process such as  ``$\chi,\chi'\to X,Y$'' which is kinematically forbidden at small relative velocity. {\tt calcSpectrumPlus} includes all  reactions with virtual {\tt X} and {\tt Y} particles. The parameter {\tt outP} specifies the spectrum: 0 - photons, 1- positrons, 2- anti-protons, 3,4,5 - neutrinos. {\tt Spect} is an array of size NZ which stores the calculated spectrum. The routine returns the  value of  $v\sigma$ in cm$^3$/s units. If the DM particles are not self-conjugated, the corresponding conjugated channel is automatically added.

\subsubsection{Direct detection}
New functions that provide the SI 90\% DD limits on light DM tabulated from the results of XENONnT~\cite{XENON:2024hup}, PandaX-4T(S2 only)~\cite{PandaX:2025rrz}, LZ~\cite{LZ:2025igz} and DarkSide50~\cite{DarkSide-50:2025lns} are accessible through\\
\noindent $\bullet$ \verb|XENON_NT(M)|  for 3~GeV $<m_{\rm DM}< 12$~GeV,\\
\noindent $\bullet$ \verb|PandaXS2(M)|  for 2~GeV $<m_{\rm DM}< 6$~GeV.\\
\noindent $\bullet$ \verb|LZ5T_light(M)|  for 3~GeV $<m_{\rm DM}< 9$~GeV.\\
\noindent $\bullet$ \verb|DS50_2025(M)|  for 0.8~GeV $<m_{\rm DM}< 5$~GeV.\\
\noindent $\bullet$ \verb|DS50_2025_noB(M)|  for 1.25~GeV $<m_{\rm DM}< 5$~GeV.\\
The DarkSide limits are given with and without (noB) binomial quenching fluctuation.

The SD 90\% DD limits on light DM from LZ5T~\cite{LZ:2025igz} are available as well\\
\noindent $\bullet$ \verb|LZ5T_lightSDp(M)|  for 3~GeV $<m_{\rm DM}< 9$~GeV.\\
\noindent $\bullet$ \verb|LZ5T_light_SDn(M)|  for 3~GeV $<m_{\rm DM}< 9$~GeV.

The projection of the future DARWIN-XLZD experiment~\cite{Baudis:2024jnk} is accessible through the function\\
\noindent $\bullet$ \verb|Darwin_XLZD(M)|  for 4.8~GeV $<m_{\rm DM}< 10$~TeV,\\
while\\
\noindent $\bullet$ \verb|Neutrino_FloorXeSI(M)| for 0.1~GeV $<m_{\rm DM}< 10$~TeV\\
corresponds to the neutrino floor for SI interactions on Xenon~\cite{OHare:2021utq}.

\subsubsection{Collider constraints}
\noindent $\bullet$ \verb|ZpLimCMS(ZpName)|\\
returns a value greater than 1 if the $Z'$ interactions with quarks and leptons are excluded at 95\% C.L. according to the CMS analysis of dilepton resonance production based on 140~fb$^{-1}$ from Run~2 data~\cite{CMS:2021ctt}. The argument \texttt{ZpName} is the name of the $Z'$ particle in the Beyond-Standard-Model (BSM) extension under investigation, as defined in the corresponding \texttt{CalcHEP} model file.

\section{Sample output} \label{sec:output}
An example of the usage of the code for the case of early matter domination is provided in the \verb|ZpPortal| model. A call to \verb|micromegas_7.1/ZpPortal/main infl1.par| will give the following output:

\begin{verbatim}
Dark matter candidate is '~dm' with spin=1/2 mass=1.00E+02

=== MASSES OF HIGGS AND ODD PARTICLES: ===
Higgs masses and widths
      h   125.00 3.82E-03

Masses of odd sector Particles:
~dm      : Mdm     = 100.000 || 

Calculation of  DM relic density
Late Inflaton decay
HInfl=2.000000E-08 GammaInfl=1.000000E-18 BrInfl=0.000000E+00
Omega=1.229275E-01 isWIMP=0
gZpDm=6.000000E-05

==== Indirect detection =======Ncdm=1 
    Channel          vcs[cm^3/s]
==================================
 annihilation cross section 8.42E-40 cm^3/s
 contribution of processes
  ~dm,~Dm -> e E                  1.62E-01
  ~dm,~Dm -> m M                  1.62E-01
  ~dm,~Dm -> l L                  1.62E-01
  ~dm,~Dm -> ne Ne                8.11E-02
  ~dm,~Dm -> nm Nm                8.11E-02
  ~dm,~Dm -> nl Nl                8.11E-02
  ~dm,~Dm -> d D                  5.41E-02
  ~dm,~Dm -> u U                  5.41E-02
  ~dm,~Dm -> s S                  5.41E-02
  ~dm,~Dm -> c C                  5.41E-02
  ~dm,~Dm -> b B                  5.41E-02
Photon flux  for angle of sight f=0.10[rad]
and spherical region described by cone with angle 0.10[rad]
Photon flux = 2.17E-26[cm^2 s GeV]^{-1} for E=50.0[GeV]
Positron flux  =  8.61E-24[cm^2 sr s GeV]^{-1} for E=50.0[GeV] 
Antiproton flux  =  7.47E-24[cm^2 sr s GeV]^{-1} for E=50.0[GeV] 

==== Calculation of CDM-nucleons amplitudes  =====
CDM[antiCDM]-nucleon micrOMEGAs amplitudes for ~dm 
proton:  SI  -3.600E-14 [3.600E-14]  SD  0.000E+00 [0.000E+00]
neutron: SI  -3.600E-14 [3.600E-14]  SD  0.000E+00 [0.000E+00]
CDM[antiCDM]-nucleon cross sections[pb]:
 proton  SI 5.560E-19 [5.560E-19] SD 0.000E+00 [0.000E+00]
 neutron SI 5.560E-19 [5.560E-19] SD 0.000E+00 [0.000E+00]

Zp :   total width=7.770014E-06 
 and Branchings:
1.536242E-01 Zp -> e,E
7.681209E-02 Zp -> ne,Ne
1.536242E-01 Zp -> m,M
7.681209E-02 Zp -> nm,Nm
1.536242E-01 Zp -> l,L
7.681209E-02 Zp -> nl,Nl
5.120805E-02 Zp -> d,D
5.120805E-02 Zp -> u,U
5.120796E-02 Zp -> s,S
5.120807E-02 Zp -> c,C
5.120807E-02 Zp -> b,B
4.656845E-02 Zp -> t,T
6.082512E-03 Zp -> ~dm,~Dm
\end{verbatim}
Note that {\tt isWIMP=0} indicates that the DM was never in equilibrium with the SM bath and behaves as a FIMP. As expected, such a particle has extremely small direct- and indirect-detection cross sections. 

\section{Conclusion} \label{sec:conclusions}
We have presented \MO\ 7, a major upgrade of the code that significantly extends its range of applicability for DM phenomenology. The central new development is the implementation of a generalized treatment of the cosmological background that enters the Boltzmann equations for DM production. This framework allows the user to go beyond the standard radiation-dominated scenario by including modified expansion histories, entropy injection, and non-thermal DM production from the decay of an additional field. It can therefore be applied to a broad class of early-Universe scenarios, including low-temperature reheating, early matter domination, kination-like cosmologies, and multi-component dark sectors.

This release also improves the treatment of light DM. In particular, \MO\ now includes a dedicated description of sub-GeV DM annihilation into light mesons through a scalar mediator. Moreover, several experimental and observational constraints have  been updated or newly implemented. 

Altogether, these developments make \MO\ 7 a unified tool for testing particle DM models against cosmological, astrophysical, direct-detection, and collider constraints in both standard and non-standard cosmological settings. 

The new version \MO\verb|_7| is available at \url{https://micromegasdm.github.io/}.

\section*{Acknowledgements}
GB and AP  thank the  Institute For Interdisciplinary Research in Science and Education (IFIRSE), Quy Nhon, Vietnam, and {\it Rencontres du Vietnam} for their warm hospitality. GB and AP thank Aytül Adigüzel and the Department of Physics, Istanbul University, Istanbul, Turkey, for their warm hospitality. GB, NB, AG, and SC acknowledge support by the Institut Pascal and the P2I axis of the Graduate School of Physics during the Paris-Saclay Astroparticle Symposium 2024 and 2025, as well as the CNRS IRP UCMN. SC acknowledges support from the UKRI Future Leader Fellowship DarkMAP (Ref. no. MR/Y034112/1). This study was conducted within the scientific program of the National Center for Physics and Mathematics, section \#5 {\it Particle Physics and Cosmology} stage 2026-2028. NB received funding from the grants PID2023-151418NB-I00 funded by MCIU/AEI/10.13039 /501100011033/ FEDER and PID2022-139841NB-I00 of MICIU/AEI/10.13039/501100011033 and FEDER, UE. AB acknowledges the use of the IRIDIS High Performance Computing Facility, and associated support services at the University of Southampton which was very important to conduct this study. AB acknowledges partial  support from the STFC grant ST/L000296/1 and  support from the Leverhulme Trust project MONDMag (RPG-2022-57). AB also thanks the NExT Institute for partial support. We thank L. Scotto Lavina for providing information on direct detection limits on light dark matter, T. Slatyer for useful discussions on CMB limits and for helping clarify the origin of the high-mass discrepancy in $f_{\rm eff}$ for quark and gluon final states, T. Sjöstrand for discussions on \textsc{Pythia}~8 and A. Roy for improving the compatibility of the code with recent macOS. We also thank A. Solomin for useful discussions on the details of simulating dark matter annihilation spectra with \textsc{Pythia}~8 and on the parallelization of the \textsc{Pythia} runs.

\appendix
\section{Appendix}\label{appendix}
To illustrate the impact of some non-standard cosmological scenarios on the DM relic density, we use  two models, the inert doublet model (IDM) where DM is a scalar  WIMP and the $Z'$ portal model (ZpPortal) where we choose parameters such that DM is a FIMP. For the IDM we choose  the parameters of the dark sector to be 
\begin{equation}
    {\rm MHX} =300~{\rm GeV}, \;{\rm MH3}= 301~{\rm GeV},\;{\rm MHC}= 304~{\rm GeV},\;\lambda_L =0.001,\;\lambda_2 = 0.01,
    \label{eq:bench}
\end{equation}
where MHX, MH3, MHC (in the notation of \MO) are the masses of the inert neutral scalar, pseudoscalar and charged scalar, respectively. For the $Z'$ model, the input parameters of the dark sector are fixed to
\begin{equation}
    M_\chi=100~{\rm GeV},\; M_{Z'}=500~{\rm GeV},\;g_{Z'}=10^{-4},\;
    \label{eq:zpbench}
\end{equation}
while the value of the coupling of the $Z'$ to DM, $g_{Z'\chi}$, varies. These parameters are defined in the files \texttt{infl1.par} in the respective directory of the IDM and ZpPortal models. 

\begin{figure}[t!]
	\centering
	\includegraphics[angle=0,height=7.cm,width=7.cm]{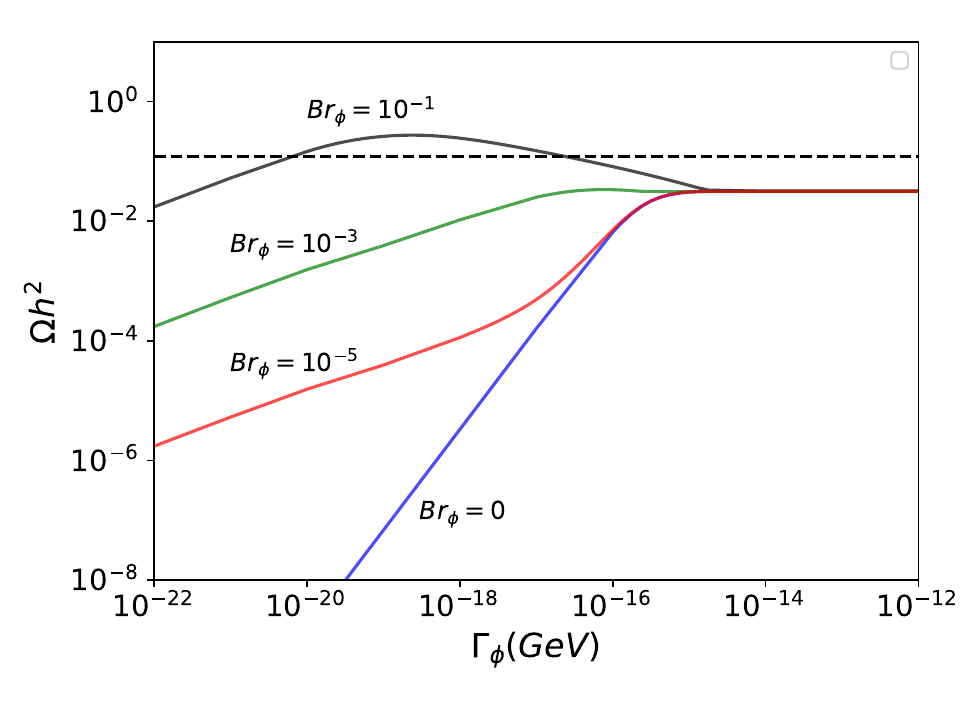}
    \includegraphics[angle=0,height=7.cm,width=7.cm]{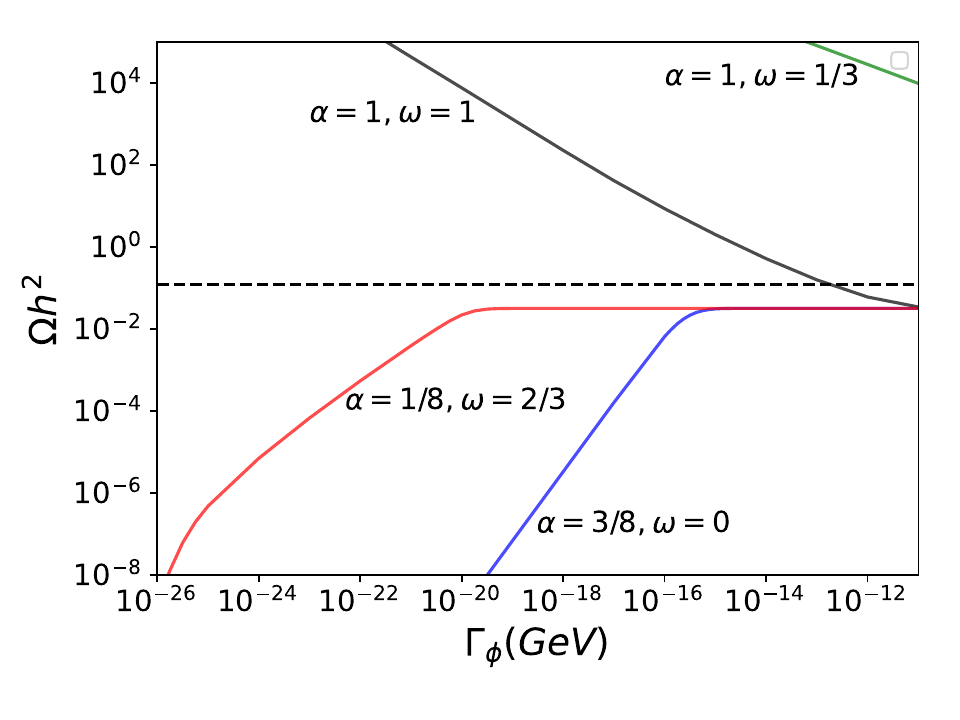}
	\caption{Relic density as a function of $\Gp$ in the IDM  for different values of the branching fraction of $\phi$ decaying  into DM (left) and  for different choices of $\alpha$ and $w$  when $\text{Br}_\phi=0$ (right). In both panels $H_\text{SM}=0$, $H_\phi=100$~GeV and the dashed lines indicate $\Omega h^2 =0.12$.}
	\label{fig:zp_br}
\end{figure}
First, we consider the early matter domination scenario with $\alpha=3/8$, $w=0$ and allow the scalar $\phi$ to decay into DM. In the case of WIMPs, this will lead to an increase of the DM relic density if $\Gp$ is small and the decay occurs after freeze-out while for large values of $\Gp$ we do not expect any impact   on the relic density. This can be seen in Fig.~\ref{fig:zp_br} (left) for the IDM benchmark in Eq.~\eqref{eq:bench}. Here we have chosen $H_\phi=100$~GeV, $H_\text{SM}=0$,  and $b=2\, \text{Br}_\phi/M_\phi$ assuming $M_\phi=10^{10}$~GeV.

The parameters of the inflaton potential also impact the DM relic density. This is illustrated in Fig.~\ref{fig:zp_br} (right) in the IDM where DM is a WIMP  for different values of $\alpha$ and $w$. As expected, the relic density is strongly suppressed at small $\Gp$ in the case of early matter domination ($\alpha=3/8$, $w=0$) and is strongly enhanced in the case of kination ($\alpha=1$, $w=1$).

\begin{figure}[t!]
	\centering
	\includegraphics[angle=0,height=6.cm,width=7.cm]{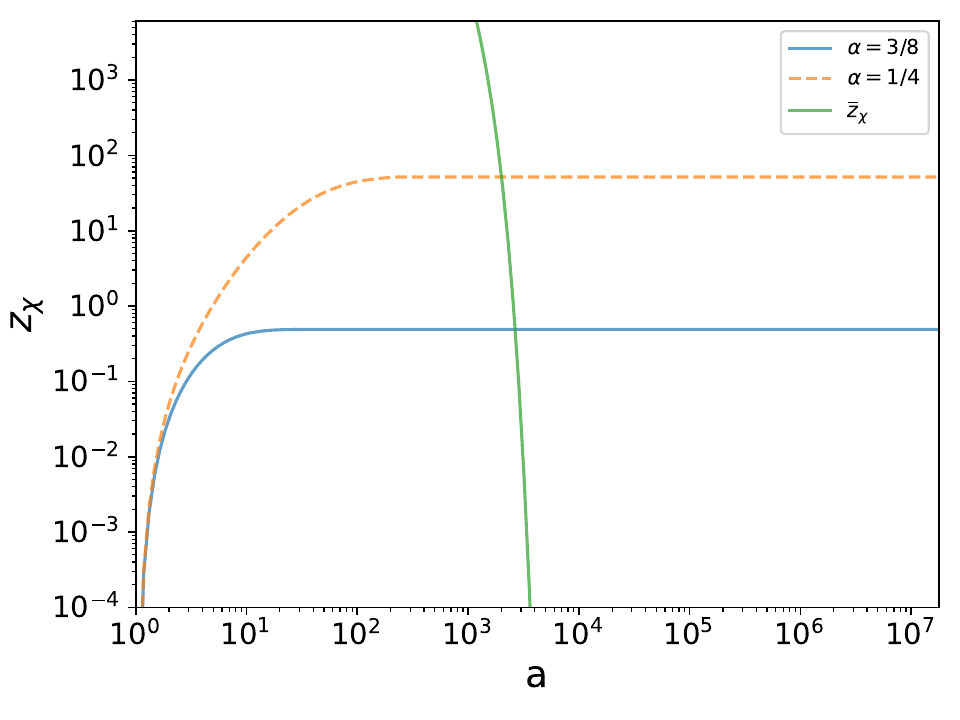}
    \includegraphics[angle=0,height=6.2cm,width=7.cm]{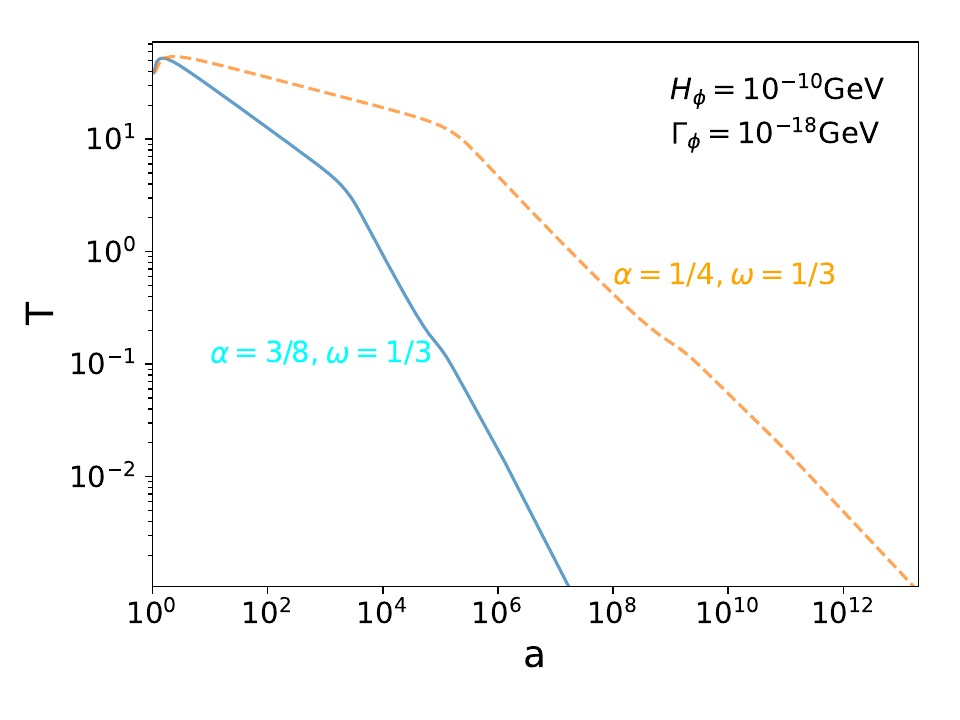}
	\caption{$z_\chi$ (left) and $T$ (right) as a function of the scale factor for two different values of $\alpha$ and $w=1/3$.  The parameters  of the $Z'$ portal model are chosen as in Eq.~\eqref{eq:zpbench} with $g_{Z'\chi}=2\times 10^{-8}$,  $H_\phi=10^{-10}$~GeV and  $\Gp=10^{-18}$~GeV.}
	\label{fig:zp_alpha}
\end{figure}
For FIMPs, we expect an increase in production if the temperature evolution is slower as the scale factor increases, that is, for a smaller value of $\alpha$. This is illustrated in Fig.~\ref{fig:zp_alpha} in  the $Z'$ portal model with a FIMP DM for  $w=1/3$ and  $\alpha=3/8 ~{\rm or}~ 1/4$ where we show both the evolution of $z_\chi$ and of $T$.  The corresponding values for the relic density are    $\Omega h^2=  5.1\times 10^{-4}$ for $\alpha=3/8$ and $\Omega h^2 = 5.3\times 10^{-2}$ for $\alpha=1/4$. 

\begin{figure}[t!]
	\centering
    \includegraphics[angle=0,height=6.cm,width=7.cm]{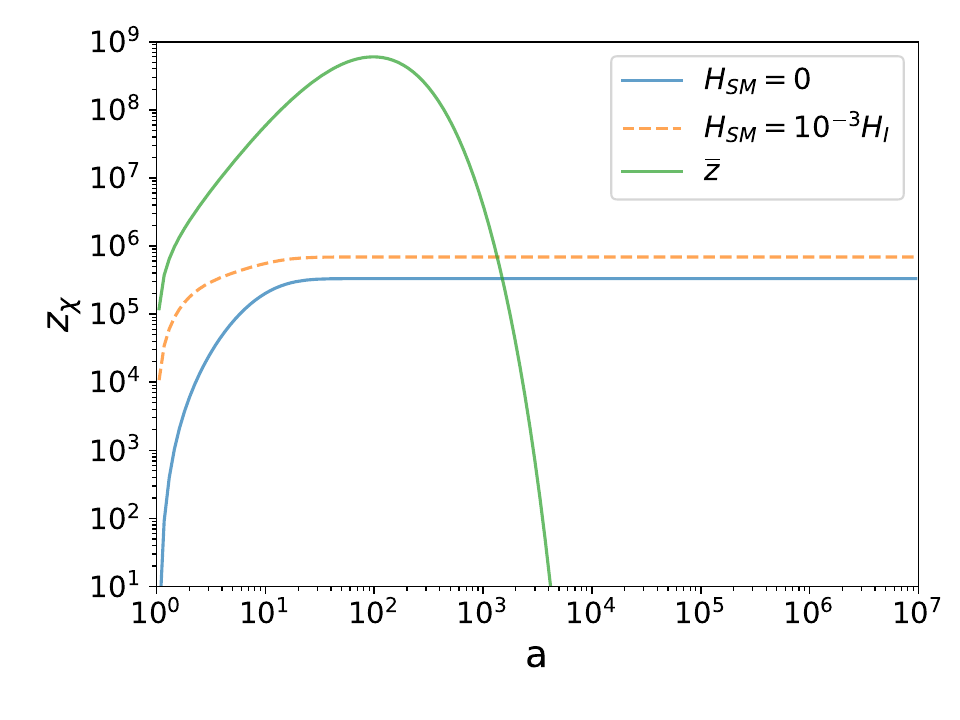}
	\caption{$z_\chi$ as a function of the scale factor for two different values of $H_\text{SM}=0$ or $H_\text{SM} =10^{-3} H_\phi$ when  $\alpha=3/8$, $w=0$.  The parameters  of the $Z'$ portal model are chosen as in Eq.~\eqref{eq:zpbench} with  $g_{Z'\chi}=6 \times 10^{-7}$, $\Gp=10^{-17}$~GeV and $H_\phi=10^{-10}$~GeV.}
	\label{fig:zp_H}
\end{figure}
Finally, we change the initial conditions by allowing $\rho_\text{SM}(t_I)$ to be larger  than $\rho_\phi$. This is parameterized by two variables $H_\phi$ and $H_\text{SM}$ in Eq.~\eqref{eq:H}. This leads to a large difference in the abundance at  high temperatures, but has little impact on the DM relic density for freeze-out as the difference  quickly disappears as the temperature decreases. An impact is only expected if the freeze-out occurs at large temperatures (for example, for DM at the TeV scale). The effect is more noticeable for the freeze-in case, as can be seen in Fig.~\ref{fig:zp_H}  for the $Z'$ portal benchmark. The corresponding shift in the relic density is $\Omega h^2= 0.103$ $(0.214)$ for $H_\text{SM}=0$ $(10^{-3}H_\phi)$.

\bibliographystyle{JHEP}
\bibliography{biblio}
\end{document}